\documentclass[prb,twocolumn,superscriptaddress,preprintnumbers]{revtex4-1}
\usepackage{graphicx}% Include figure files
\usepackage{dcolumn}% Align table columns on decimal point
\usepackage{bm}% bold math
\usepackage{hyperref}% add hypertext capabilities
\usepackage{subfigure}
\usepackage{wrapfig}
\usepackage{amssymb,latexsym, amsmath}
\usepackage{graphics}
\usepackage{float}
\usepackage{appendix}
\usepackage{amssymb,latexsym,amsthm,makeidx,enumerate}
\usepackage{mathrsfs}

\usepackage{siunitx}
\usepackage{comment,hhline,multirow,setspace}

\begin{document}

\makeatletter
\let\NAT@bare@aux\NAT@bare
\def\NAT@bare#1(#2){%
	\begingroup\edef\x{\endgroup
		\unexpanded{\NAT@bare@aux#1}(\@firstofone#2)}\x}
\makeatother

\title{Fermi surface evolution of Na-doped PbTe studied through density functional theory calculations and Shubnikov-de Haas measurements}

\author{P. \surname{Giraldo-Gallo}}
\affiliation{Geballe Laboratory for Advanced Materials, Stanford University, Stanford, CA 94305, USA}
\affiliation{Department of Physics, Stanford University, CA 94305, USA}
\affiliation{National High Magnetic Field Laboratory, Tallahassee, Florida 32310, USA}
\author{B. Sangiorgio}
\affiliation{Materials Theory, ETH Zurich, Wolfgang-Pauli-Strasse 27, CH-8093 Z\"urich, Switzerland}
\author{P. Walmsley}
\affiliation{Geballe Laboratory for Advanced Materials, Stanford University, Stanford, CA 94305, USA}
\affiliation{Department of Applied Physics, Stanford University, CA 94305, USA}
\author{M. Fechner}
\affiliation{Materials Theory, ETH Zurich, Wolfgang-Pauli-Strasse 27, CH-8093 Z\"urich, Switzerland}
\author{S. C. Riggs}
\affiliation{National High Magnetic Field Laboratory, Tallahassee, Florida 32310, USA}
\author{T. H. Geballe}
\affiliation{Geballe Laboratory for Advanced Materials, Stanford University, Stanford, CA 94305, USA}
\affiliation{Department of Applied Physics, Stanford University, CA 94305, USA}
\author{N. A. Spaldin}
\affiliation{Materials Theory, ETH Zurich, Wolfgang-Pauli-Strasse 27, CH-8093 Z\"urich, Switzerland}
\author{I. R. Fisher}
\affiliation{Geballe Laboratory for Advanced Materials, Stanford University, Stanford, CA 94305, USA}
\affiliation{Department of Applied Physics, Stanford University, CA 94305, USA}

\date{\today}

\begin{abstract}

We present a combined experimental and theoretical study of the evolution of the low-temperature Fermi surface of lead telluride, PbTe, when holes are introduced through sodium substitution on the lead site. Our Shubnikov-de-Haas measurements for samples with carrier concentrations up to $9.4\times10^{19}$cm$^{-3}$ (0.62 Na atomic $\%$) show the qualitative features of the Fermi surface evolution (topology and effective mass) predicted by our density functional (DFT) calculations within the generalized gradient approximation (GGA): we obtain perfect ellipsoidal L-pockets at low and intermediate carrier concentrations, evolution away from ideal ellipsoidicity for the highest doping studied, and cyclotron effective masses increasing monotonically with doping level, implying deviations from perfect parabolicity throughout the whole band. Our measurements show, however, that standard DFT calculations underestimate the energy difference between the L-point and $\Sigma$-line valence band maxima, since our data are consistent with occupation of a single Fermi surface pocket over the entire doping range studied, whereas the calculations predict an occupation of the $\Sigma$-pockets at higher doping. Our results for low and intermediate compositions are consistent with a non-parabolic Kane-model dispersion, in which the L-pockets are ellipsoids of fixed anisotropy throughout the band, but the effective masses depend strongly on Fermi energy.

\end{abstract}

%\pacs{74.81.-g,74.62.En,74.40.Kb}
%\keywords{}

\maketitle

\section{Introduction}

Lead telluride, PbTe, is a widely known thermoelectric material and a narrow-gap semiconductor, which can be degenerately doped by either Pb (hole-doping) or Te (electron-doping) vacancies, or by introduction of acceptor or donor impurities \cite{Ravichbook2, Dornbook, Khokhlov}. Such impurity dopants have been shown to enhance the thermoelectric figure of merit, $zT$, from 0.8 to 1.4 for the case of sodium doping \cite{Snyder2,Snyder1,Kanatzidis1}, and to 1.5 for doping with thallium \cite{Snyder2,Snyder3}. Tl is also the only dopant known to date that leads to a superconducting ground state in PbTe; remarkably its maximum critical temperature of $T_c$=1.5 K is almost an order of magnitude higher than other superconducting semiconductors with similar carrier density \cite{Ravich,Lewis1,Yana1,Yana2,DZero}. Understanding the physical origin of these enhanced properties and their dependence on the choice of dopant chemistry requires a detailed knowledge of the electronic structure, in particular its evolution with changes in dopant and carrier concentrations.

\begin{figure}[!t]
%\hspace{-0.5cm}
\centering
\includegraphics[scale=0.45]{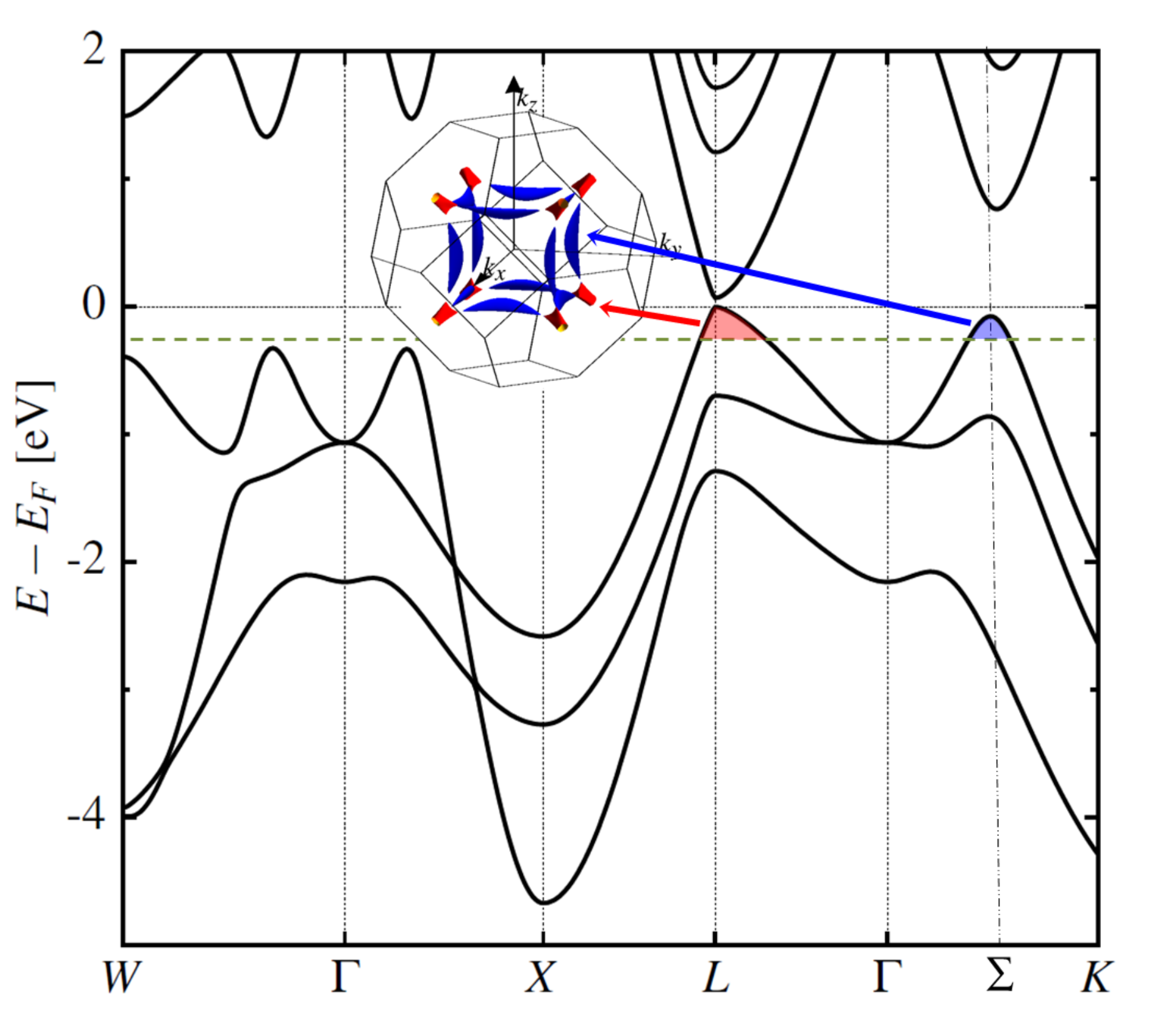} %\hspace{-0.5cm}
\caption{(Color online) Energy dispersion for stoichiometric PbTe along the high symmetry directions of the FCC Brillouin zone, calculated in this work using density functional theory (DFT) (for details see text). A direct gap, underestimated compared with experiment as is usual in DFT calculations, is observed at the L point, and a second valence band maximum occurs along the $\Sigma$ high-symmetry line. A representative Fermi surface, which emerges as the Fermi energy is shifted into the valence band by Pb vacancies or hole-dopant impurities, is shown in the inset. For the choice of Fermi level shown (green-dashed line), the Fermi surface contains eight half-ellipsoids (shaded in red) centered at the L-point and oriented along the [111] directions (L-pockets), and twelve $\Sigma$-pockets (shaded in blue) centered closed to the mid-point of the [110] $\Sigma$ line and oriented along the [100] directions.}\label{fig_Evsk}
\end{figure}

\begin{figure*}[!t]
%\hspace{-0.5cm}
\centering
\includegraphics[scale=0.45]{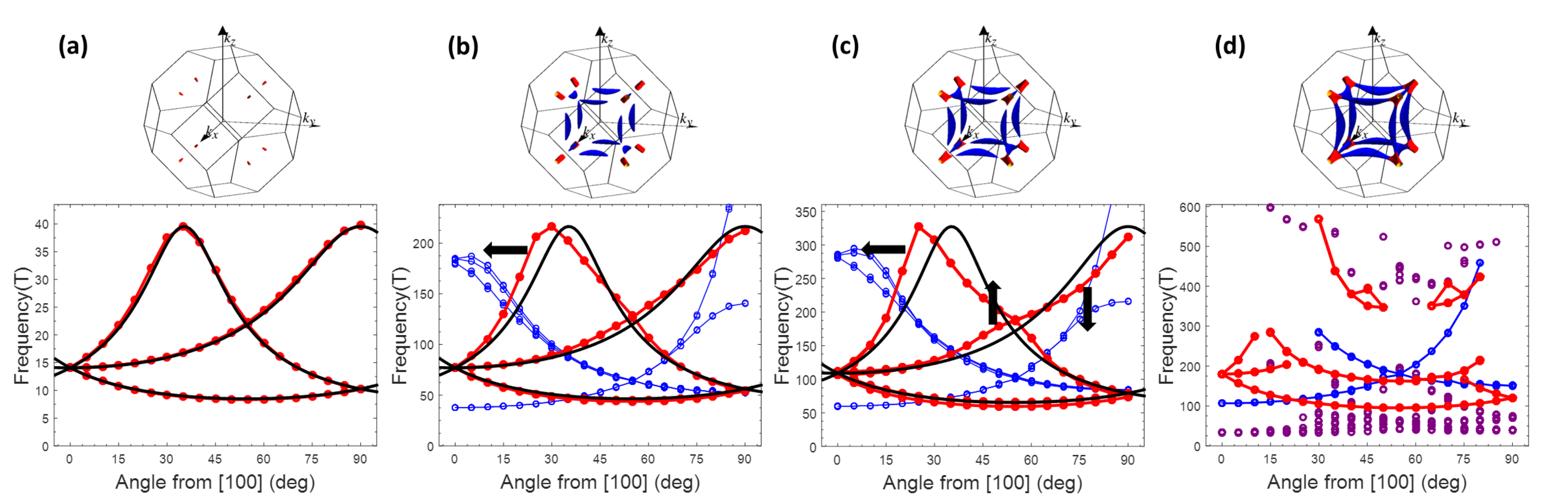}
\caption{(Color online) Upper panel: Fermi surface of hole-doped PbTe calculated in this work using the rigid band approximation. Lower panel plots: The corresponding (110)-plane angle evolution of the cross-sectional areas (in frequency units) of the calculated Fermi surface pockets. The four columns correspond to monovalent impurity concentrations of: {\bf (a)} $x=0.02\%$ ($p_L=0.27\times 10^{19}$cm$^{-3}$ and $p_{\Sigma}=0$); {\bf (b)} $x=0.81\%$ ($p_L=3.5\times 10^{19}$cm$^{-3}$ and $p_{\Sigma}=8.6\times 10^{19}$cm$^{-3}$); {\bf (c)} $x=1.56\%$ ($p_L=6.1\times 10^{19}$cm$^{-3}$ and $p_{\Sigma}=17.4\times 10^{19}$cm$^{-3}$); and {\bf (d)} $x=2.61\%$ ($p_{total}=36.2\times 10^{19}$cm$^{-3}$). The frequencies of the L-pockets are shown in red, and compared with those expected in a perfect ellipsoidal model shown as black lines. The evolution of the $\Sigma$-pockets is shown in blue. These pockets appear at a dopant concentration of $x=0.11\%$ ($p_L\approx 10^{19}$cm$^{-3}$). In column (d), the $\Sigma$ and L-pockets have merged, forming a cube-shape Fermi surface; cross-sections that can not be identified separately with $\Sigma$ or L are shown in purple. We plot frequencies up to 600 T, noting, however, that frequencies up to 8 kT occur, corresponding to the large-square Fermi surface orbits.}\label{fig_boris}
\end{figure*}

The valence band of PbTe has two maxima, located at the L point and close to the mid-point of the $\Sigma$ high-symmetry line (we call this the $\Sigma_m$ point) of the Brillouin Zone (see Figure 1). The enhancement of $zT$ with doping has been recently suggested to be at least in part associated with a decrease in the effective dimensionality of parts of the Fermi surface as the $\Sigma_m$ pockets connect (Figure~\ref{fig_boris}) \cite{Parker}. For the case of superconductivity, an increase of the density of states at the Tl concentration for which superconductivity emerges, as a consequence of the appearance of an additional band, has been invoked as a possible explanation for the enhanced $T_c$ \cite{Lewis1}. Such hypotheses can be tested by a direct experimental determination of the Fermi surface topology and its evolution with carrier concentration. To date, such studies have been limited to quantum oscillation measurements performed in the low carrier concentration regime ($p\leq 1.1\times 10^{19}$cm$^{-3}$ for full topology) \cite{burke2, burke1}, although the enhanced thermoelectric and superconducting properties occur at considerably higher carrier concentrations. A direct measurement of the Fermi surface characteristics for these higher carrier densities is clearly needed. 

In this paper we present the results of a detailed computational and experimental study of the fermiology of $p$-type Na-doped PbTe (Pb$_{1-x}$Na$_x$Te), with carrier concentrations up to $9.4\times 10^{19}$cm$^{-3}$, obtained via density functional theory (DFT) calculations of the electronic structure, and measurements of quantum oscillations in magnetoresistance for fields up to 35 T. These measurements enable a direct characterization of the Fermi surface morphology and quasiparticle effective mass for values of the Fermi energy that far exceed those available by self-doping from Pb vacancies. Our main findings are: \\
(i) At low temperatures, the Fermi surface is formed from eight half ellipsoids at the L points (the L-pockets) with their
primary axes elongated along the [111] directions. The Fermi surface is derived from a single band up to the
highest carrier concentration measured, $9.4\times 10^{19}$cm$^{-3}$.\\
(ii) The L-pockets are well described by a perfect ellipsoidal model up to a carrier concentration of $6.3\times 10^{19}$cm$^{-3}$. For a carrier concentration of $9.4\times 10^{19}$cm$^{-3}$, subtle deviations from perfect ellipsoidicity can be resolved. These deviations are qualitatively consistent with those predicted by the band structure calculations.\\
(iii) The effective cyclotron masses increase monotonically with carrier concentration for all high-symmetry directions, implying that the L band is not well described by a perfect parabolic model for any carrier density. This evolution is also consistent with the predictions from our band structure calculations.\\
(iv) Although the qualitative evolution of the Fermi surface topology with carrier concentration is correctly predicted by band structure calculations, these calculations underestimate the band-offset (between the top of the L-band and the top of the $\Sigma_m$-band).  

Before detailing our experiments, we emphasize that our measurements are made in the low temperature regime and caution should be exercised before extrapolating the results to different temperature regimes. Quantum oscillations characterize the low-temperature properties of a material, and due to the exponential damping factor, they cannot be observed above approximately 60 K in Na-doped PbTe. Hence, we do not claim that our first three findings outlined above necessarily remain valid at higher temperatures. In particular, earlier experimental studies, based on magnetoresistance and Hall coefficient measurements \cite{Sitter1}, have indicated an appreciable temperature dependence of both the band gap and the band offset (between L and $\Sigma$ band maxima) in PbTe. The current measurements provide a definitive determination of the morphology of the Fermi surface at low temperatures, and hence provide an important point of comparison for band structure calculations, but additional measurements based on a technique that is less sensitive to the quasiparticle relaxation rate, such as angle resolved photo emission Spectroscopy (ARPES), are required in order to determine whether the $\Sigma$-pocket remains unoccupied at higher temperatures.

\section{First-principles Calculations}

To provide a baseline with which to compare our experimental data, we first perform density functional theory (DFT) 
calculations of the electronic structure of PbTe with and without doping. 
An accurate description of this compound within DFT is very challenging; in particular the 
computed properties are highly sensitive to the choice of volume (as already reported in Refs. \citenum{an2008,zhang2009}), the exchange-correlation functional, and whether or not spin-orbit coupling is included. 
A change in lattice constant of $1\%$, for example, can both change the band offset by $60\%$ and generate a ferroelectric instability. Moreover, when spin-orbit coupling is included, an unusually fine $k$-point mesh is needed to converge the phonon frequencies, forces and Fermi energy. This unusual sensitivity to the input parameters in the calculation is of course
related to the many interesting properties of PbTe, which is on the boundary between various competing structural (incipient ferroelectricity\cite{bozin2010,Ann2}) 
and electronic (superconductivity \cite{Yana1,DZero,Yana2} and topological insulator\cite{barone2013,barone2013a}) instabilities.

\subsection{Computational details}

Our calculations were performed using the PAW implementation \cite{bloechl1994,kresse1999} of density functional theory
within the VASP package \cite{kresse1996}. After carefully comparing structural and electronic properties calculated 
using the local density approximation (LDA) \cite{perdew1981}, PBE \cite{perdew1996} and PBEsol \cite{perdew2008} with 
available experimental data, we chose the PBEsol exchange-correlation functional as providing the best overall agreement.  
We used a $20\times20\times20$ $\Gamma$-centered $k$-point mesh and to ensure a convergence below $0.1$ $\mu$eV for the total energy used
a plane-wave energy cutoff of $600$ eV and an energy threshold for the self-consistent calculations of 0.1 $\mu$eV. We
used valence electron configurations $5d^{10}6s^26p^2$ for lead, $5s^2 5p^4$ for tellurium, and $2p^63s^1$ for sodium. Spin-orbit coupling was 
included. The unit cell volume was obtained using a full structural relaxation giving a lattice constant of 6.44 \AA (to be compared with the experimental 6.43 \AA\cite{hummer2007}). Kohn-Sham band energies were computed on a fine ($140\times140\times140$) three-dimensional grid covering the entire Brillouin zone and used as an input for the SKEAF code \cite{rourke2012} which allows for extraction of extremal cross-sectional areas of the Fermi surface in different spatial orientations. 

\subsection{Rigid-band approximation}

First, we computed the Fermi-surface evolution as a function of doping (shown in Fig.~\ref{fig_boris}) by rigidly shifting the Fermi energy in the pure PbTe structure and assuming one hole per dopant. This rigid-band approximation allows very fine samplings of the Brillouin zone, which are necessary to characterize the tiny Fermi surface of hole-doped PbTe at low doping. We discuss its validity here, by comparing with calculations in which a Pb ion is substituted explicitly with a Na ion.
Many first-principles studies\cite{ahmad2006,hase2006,Salameh,Cho1,takagiwa2013,lee2012} have already been carried out to determine the effect of different dopant atoms on the electronic properties of PbTe, with  
some of them explicitly assessing the validity of the rigid band approximation in Na-doped PbTe: \citet{takagiwa2013} confirmed from KKR-CPA calculations that the density of states (DOS) behaves as in a rigid band model, whereas \citet{hoang2010} and \citet{lee2012} showed that a lifting of degeneracy occurs at the top of the valence band with explicit Na doping (at a concentration of $3.125\%$), with the consequence that the rigid band approximation overestimates the thermopower\cite{lee2012}.
Here we study how sodium impurities affect the band structure of PbTe close to the Fermi energy for the lower concentrations that we use in our experiments ($x\lesssim$1$\%$). 

We show here results for a $4\times4\times4$ supercell of the primitive cell containing 128 atoms ($x\approx1.6\%$), with one lead ion substituted by sodium. The unit cell volume was kept the same as in pristine PbTe (it would be changed by less than $0.1\%$ by a full structural relaxation). We checked also that our conclusions are qualitatively unchanged for a larger 216 atom supercell ($3\times3\times3$ the conventional cubic cell) in which one or two lead ions are substituted by sodium ($x\approx0.9\%$ or $x\approx1.6\%$). 
 The $k$-point mesh was accordingly scaled down and spin-orbit coupling was not included because of computational cost; the other computational settings were left unchanged. 

\begin{figure}[!t]
	\centering
	\includegraphics[width=0.48\textwidth]{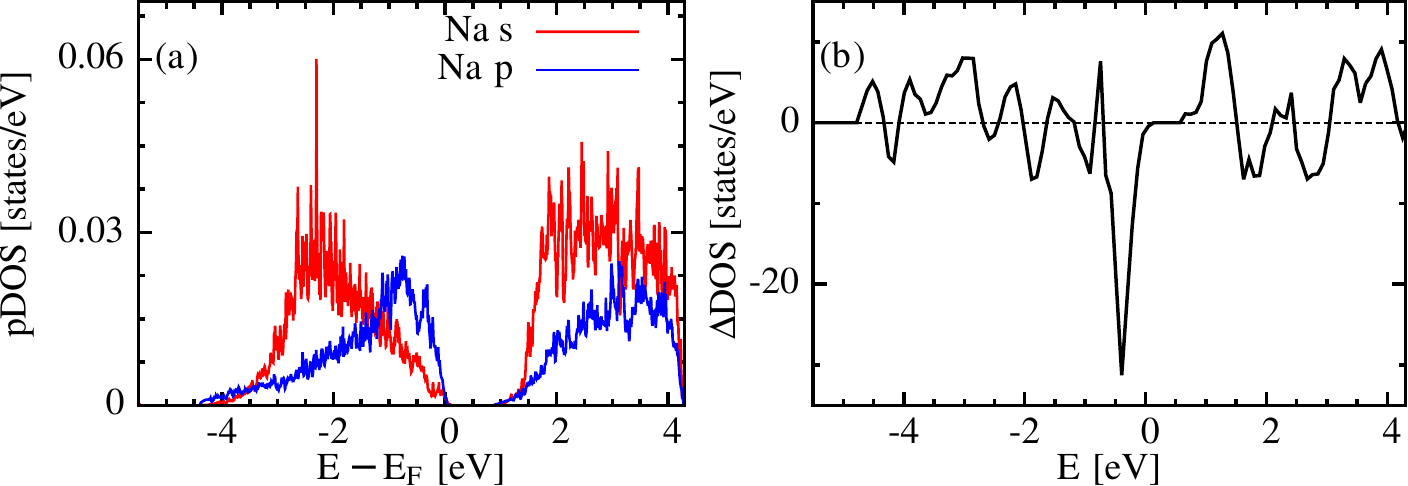} 
	\caption{(Color online) Sodium contribution to the band structure around the Fermi energy for the 128-atom supercell. (a) Sodium projected density of states (pDOS). (b) Difference in the total DOS with and without the impurity, $\Delta\mathrm{DOS}=\mathrm{DOS}_{\mathrm{with\;Na}}- \mathrm{DOS}_\mathrm{undoped}$. Note the drop in DOS just below the Fermi energy, consistent with a lifting in degeneracy of the highest valence bands (see also Figure \ref{fig_Na_bands}).}\label{fig_Na_contrib}
	%\vspace{-0.4cm}
\end{figure}

Figure~\ref{fig_Na_contrib} (a) shows the partial density of states in the region of the Fermi level (set to 0 eV) from the sodium impurity for $x\approx1.6\%$. Note
the small value on the $y$ axis indicating that the contribution from the Na atom is very small. It does, however, have an influence on the electronic band structure which 
can be seen in Figure~\ref{fig_Na_contrib} (b), where we plot the {\it difference} in density of states with and without the impurity. Here we see a distinct drop
in the DOS (note the higher values on the $y$ axis) just below the Fermi energy due to band shifts caused by the presence of the Na atom; we analyze these next.  

\begin{figure}[!t]
	\centering
	\includegraphics[width=0.48\textwidth]{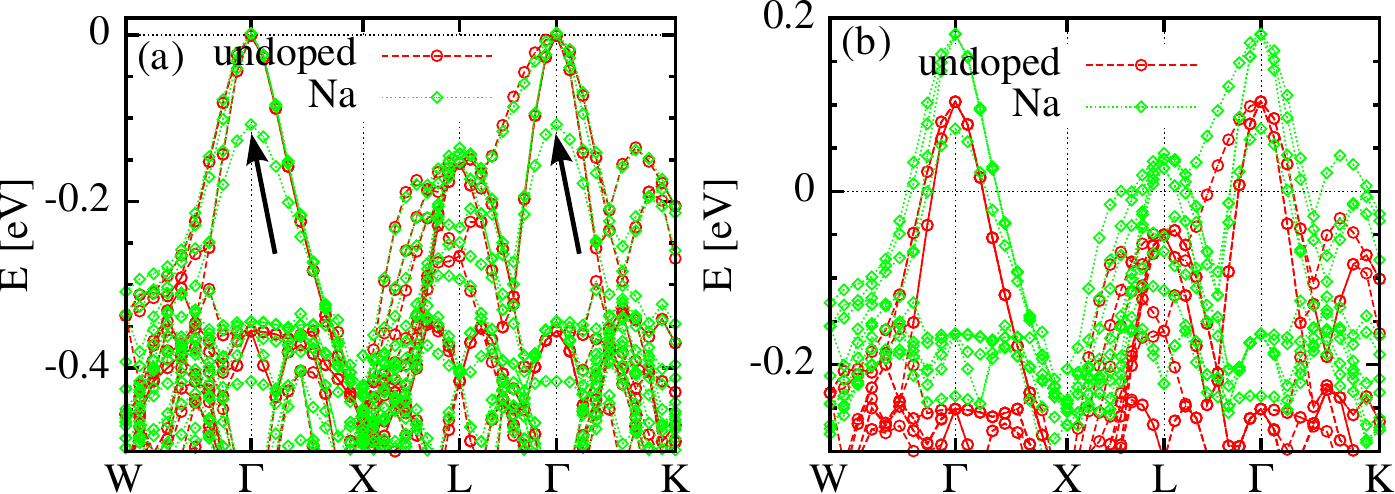} 
	\caption{(Color online) Calculated band structure with and without sodium impurity for the 128-atom supercell ($x\approx1.6\%$). {\bf (a)} The zero of energy was set at the top of the valence band for both cases. Note the lifting of the degeneracy of the top valence bands (marked by arrows); apart from this, the bands coincide almost perfectly. {\bf (b)} The carrier density for both cases was fixed to a concentration corresponding to $x=1.6\%$. The Fermi energy is moved more into the valence band than expected from the rigid band approximation because of the lifting of degeneracy. 
}\label{fig_Na_bands}
	%\vspace{-0.4cm}
\end{figure}

In Figure~\ref{fig_Na_bands} we compare the calculated electronic band structure with and without the sodium impurity. In Fig.~\ref{fig_Na_bands} (a) we show both
band structures on the same $y$ axis with the zero of energy set to the top of the valence band. We see that the two band structures are close to identical, except
for a lifting of the eight-fold degeneracy at the top of the valence band, indicated by black arrows, in the case of the explicit Na doping. A consequence of this
shift in one of the valence bands is a shift of the Fermi energy to lower energy relative to its position in the rigid band approximation. We illustrate this in
Fig.~\ref{fig_Na_bands} (b) where we set the zero of energy to be the Fermi energy for each case. 
In contrast with earlier calculations at a larger doping \cite{hoang2010,lee2012}, the lifted band \textit{does} contribute to the Fermi surface and affects
the \textit{quantitative} evolution of dHvA frequencies with hole density, giving rise to a more complex Fermi surface having L-pockets with different sizes. 
The folding of wave vectors and states in the supercell makes an estimation of the different ellipsoidal axes difficult. In any case, the amplitude of the quantum oscillations for the ``lifted-degenerate'' pockets would be weaker. 
From these considerations we are confident that our rigid-band calculations can be used to make \textit{qualitative} predictions about the evolution of the Fermi 
surface with Na doping. Quantitative predictions are anyway difficult because of the previously discussed sensitivity on the parameters used for the calculations.

\subsection{Calculated Fermi surface evolution and angle evolution of Shubnikov-de Haas frequencies}\label{sec_DFTSdH}

Our calculated energy dispersion for PbTe, along the high symmetry directions of the FCC Brillouin zone, is plotted in Figure \ref{fig_Evsk}. As discussed above, we obtain a direct gap at the L-point, followed by a second valence band maximum at the $\Sigma_m$-point, 70 meV below the top of the valence band. 
Figure~\ref{fig_boris} shows our calculated Fermi surfaces, as well as the (110)-plane angle dependence of the Fermi surface pocket cross-sectional areas, or equivalently, Shubnikov-de Haas (SdH) frequencies (see appendix \ref{app_QO}), for four impurity concentrations. The (110) plane is a natural plane to study the angle evolution of the SdH frequencies for this material, given that, in a perfect ellipsoidal scenario, it allows the determination  all the extremal cross-sectional areas of both, L- and $\Sigma$-pockets.  
For low impurity concentrations, the Fermi surface is formed only by L-pockets, which follow the angle dependence expected for a perfect ellipsoidal model. At intermediate concentrations, the $\Sigma$-pockets appear, and clear deviations from the perfect ellipsoidal model for L-pockets (and $\Sigma$-pockets) are observed. 
For impurity concentrations above $x=1.8\%$, $\Sigma$- and L-pockets merge together to form the Fermi surface shown in Figure \ref{fig_boris}(d). At this point, very high frequency ($\approx$ 8 kT, corresponding to the large-square Fermi surface pieces) and very low frequency features are expected, and a whole new variety of cross-sectional areas coming from different sections of the Fermi surface make the tracking of continuous angle dependence curves more challenging. 

\begin{figure*}[!t]
	\centering
	\includegraphics[width=\textwidth]{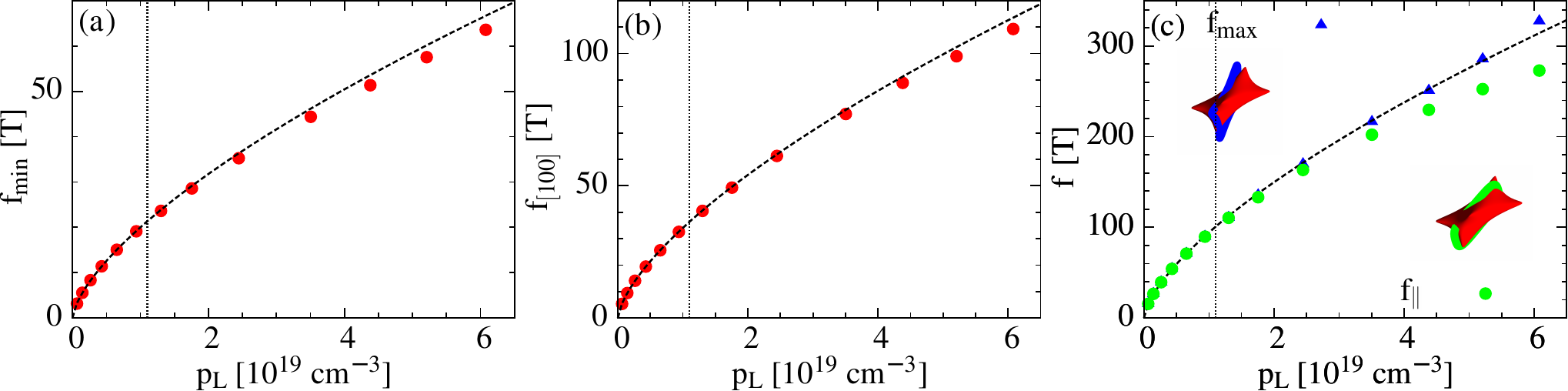}
	\caption{(Color online) Evolution of three cross-sectional areas (in frequency units) with density of holes in the L-pockets ($p_L$). The dashed curve in all the plots shows the functional dependence of $p_L^{2/3}$ expected for a perfect ellipsoidal model. The dotted vertical line indicates the L-pocket hole density above which the $\Sigma$-pockets start to be populated. {\bf (a)} Frequency associated with the L-pockets minimum cross-sectional area, $f_{min}$; {\bf (b)} Frequency associated with the L-pockets' cross-sectional area in the [100] direction, $f_{[100]}$; {\bf (c)} Frequencies associated with the L-pockets' maximum cross-sectional area. The green circles correspond to the orbits in the longitudinal direction of the L-pocket ($f_\parallel$) -- for perfect ellipsoidal L-pockets they would correspond to the largest possible frequencies; the blue triangles correspond to the orbits associated with the largest cross-sectional area $f_\mathrm{max}$, which for large concentrations do not correspond anymore to longitudinal orbits on the L-pockets. The inset shows two representative orbits ($f_\parallel$ in green and $f_\mathrm{max}$ in blue) on the distorted L-pocket (shown in red) for a concentration $x=1.56\%$ ($p_L=6\times 10^{19}$ cm$^{-3}$).}\label{fig_freq_nl_dft}
\end{figure*} 

\begin{figure}[!t]
	%\hspace{-0.5cm}
	\centering
	\includegraphics[scale=0.9]{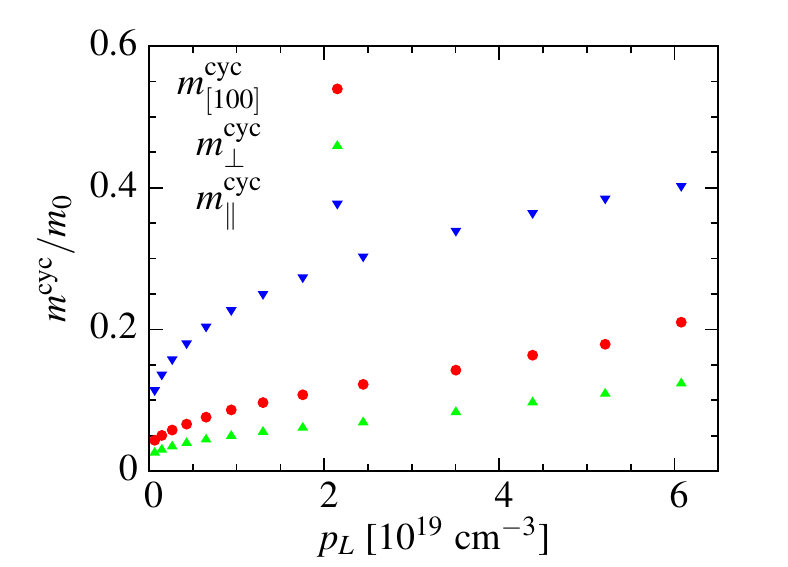} %\hspace{-0.5cm}
	\caption{(Color online) Evolution of cyclotron effective masses (Eq. \ref{eq_cyc_masses}) as a function of density of holes in the L-pocket ($p_L$) at three high symmetry directions: $\parallel$ or in the longitudinal direction of the L-pocket, in the [100] direction, and $\perp$ or in the transverse direction of the L-pocket (corresponding to a magnetic field oriented along the [111] direction). The variation with $p_L$ provides striking evidence for the non-parabolicity of the bands. }\label{fig_mstar_dft}
\end{figure}

For the L-pockets, we observe a progressive evolution to non-ellipsoidicity, characterized by three main features in the angle dependence plots: (i) an increasing splitting in the low frequency branch, indicative of deformations of the L-pockets around the minor semiaxis region; (ii) a shifting to lower values of the angle at which the maximum cross-sectional area (maximum frequency) is found, indicative of L-pocket deformations around the major-semiaxis region, and due to the formation of the tips that will eventually join with the $\Sigma$-pockets at high enough dopant concentration; (iii) some distortions of the dispersion branch that goes from the [100] frequency value to the maximum frequency value at 90$^{\circ}$, generating a cusp at 90$^{\circ}$.

Figure \ref{fig_freq_nl_dft} shows our calculation of three extremal cross-sectional areas with density of holes in the L-pockets ($p_L$) computed from the Kohn-Sham band energies. The dashed curves indicate the expected $p_L^{2/3}$ behavior for perfect ellipsoidal pockets. Deviations of the computed cross-sectional areas from the perfect ellipsoidal dependence become noticeable close to hole densities in the L-pockets above which the $\Sigma$-pockets start to be populated, which is indicated by the vertical dotted lines in Fig. \ref{fig_freq_nl_dft}. 
These deviations are characterized by a shift toward lower frequencies from that expected in the perfect ellipsoidal model. Additionally, Figure \ref{fig_freq_nl_dft}(c) highlights the distortions in the L-pockets, which among others cause the shift in the maximum frequency from $35^\circ$ ($f_\parallel$) towards smaller angles in the angle-evolution curves showed in Figures \ref{fig_boris}(b) and \ref{fig_boris}(c). 
 Note that larger band offsets -- obtained by changing the unit cell volume -- would not greatly affect these considerations, in particular the density of holes at which the $\Sigma$-pockets appear. 

Figure \ref{fig_mstar_dft} shows our calculated evolution of cyclotron effective masses (Eq. \ref{eq_cyc_masses}) at three high symmetry directions as a function of the carrier content of the L-pockets. A monotonic increase of cyclotron masses with carrier concentration is observed, implying a non-parabolicity of the L-band even at the top of the band. It is interesting to note that although deviations from perfect ellipsoidicity as seen in the calculated angle evolution (Figure \ref{fig_boris}(a)) and the calculated dHvA frequencies (Figure \ref{fig_freq_nl_dft}) are close to zero for the low carrier concentration regime, the variation of the effective masses at the lowest doping levels already points to the non-parabolicity of the highest valence bands. Note that this was already taken into account in some transport studies of PbTe to compute its thermoelectric properties \cite{venkatapathi2014,bilc2006}.

In summary, our density functional calculations of the evolution of the Fermi surface of PbTe with doping provide some guidelines for identifying signatures of deviations from perfect ellipsoidicity and perfect parabolicity in our quantum oscillation experiments, to be presented in the coming sections. As we mentioned previously, the main signatures in the angular dependence of cross-sectional areas of L-pockets are:\\
(i) An increasing splitting in the low frequency branch, indicative of deformations of the L-pockets around the minor semiaxis region; \\
(ii) A shifting to lower values of the angle at which the maximum cross-sectional area (maximum frequency) is found, indicative of L-pocket deformations around the major-semiaxis region, and due to the formation of the tips that will eventually join with the $\Sigma$-pockets at high enough dopant concentration; \\
(iii) Some distortions of the dispersion branch that goes from the [100] frequency value to the maximum frequency value at 90$^{\circ}$, generating a cusp at 90$^{\circ}$. These guidelines will be used in determining deviations from perfect ellipsoidicity in the data;\\
(iv) A monotonic increase of the cyclotron effective mass of holes as a function of carrier concentration. 

Our computational findings will be used next in interpreting deviations from perfect ellipsoidicity in our experimental data.

%\FloatBarrier

\section{Experimental Techniques}

\subsection{Sample preparation}

Pb$_{1-x}$Na$_x$Te single crystals were grown by an unseeded physical vapor transport (VT) method, similar to that described in ref. \citenum{Yana2}, by sealing in vacuum polycrystalline pieces of the already doped compound, with (or close to) the desired final stoichiometry. The polycrystalline material was obtained by mixing high purity metallic lead, tellurium and sodium in the appropriate ratios. The source materials were placed in alumina crucibles, sealed in evacuated quartz tubes, and heated up to 1000 $^{\circ}$C, holding this temperature for 7 hours, followed by a rapid quench in water. A subsequent sinter at 700 $^{\circ}$C for 48 hours was performed with the material contained in the same evacuated tube \cite{Yamini1}. After this process, the material was removed from the crucible, ground into fine powders, and then cold-pressed into a pellet. The pellet was sealed in a quartz tube, with a small argon pressure to prevent mass transport. The pellet was then sintered again at 500 $^{\circ}$C for 24 hours, and finally it was broken into small pieces to be used in the VT stage. After the VT, mm-sized single crystals, with clear cubic facets, were obtained. The final sodium content was estimated through the determination of the carrier concentration via Hall coefficient ($p_H$) measurements, assuming one hole per Na dopant. Direct determination of the dopant concentration is challenging for the low Na concentrations studied in this work ($<0.62\%$) which are below the weight $\%$ resolution of the available electron microprobe analysis tools.

\subsection{Magnetoresistance measurements}

High-field magnetoresistance measurements of Pb$_{1-x}$Na$_x$Te single crystal samples with different $x$ values between 0 and 0.62$\%$ (carrier concentrations up to $p_H=9.4\times 10^{19}$cm$^{-3}$) were taken at the DC facility of the National High Magnetic Field Laboratory (NHMFL), in Tallahassee, FL, USA, for magnetic fields up to 35 T. Pb$_{1-x}$Na$_x$Te single crystals were cleaved in rectangular shapes with faces along the [100] directions. Typical sizes of the resulting crystals were 1 mm in the longest side. Four gold pads were evaporated on one of the faces in order to improve electrical contact with the crystal. Gold wires were attached to each of the pads using silver epoxy, and the other end of each wire was pasted to a glass slide. Twisted pairs coming from the 8-pin dip socket were connected to the glass slide, with special care taken to minimize the loop areas of the wires. Four-point resistance curves for different field orientations and temperatures were taken for plus and minus field sweeps (in order to extract the symmetric component of the magnetoresistance) with temperature and field orientation held constant. 

\begin{figure*}[!htbp]
\vspace{0.3cm}
\hspace{-0.3cm}
\centering
\includegraphics[scale=0.52]{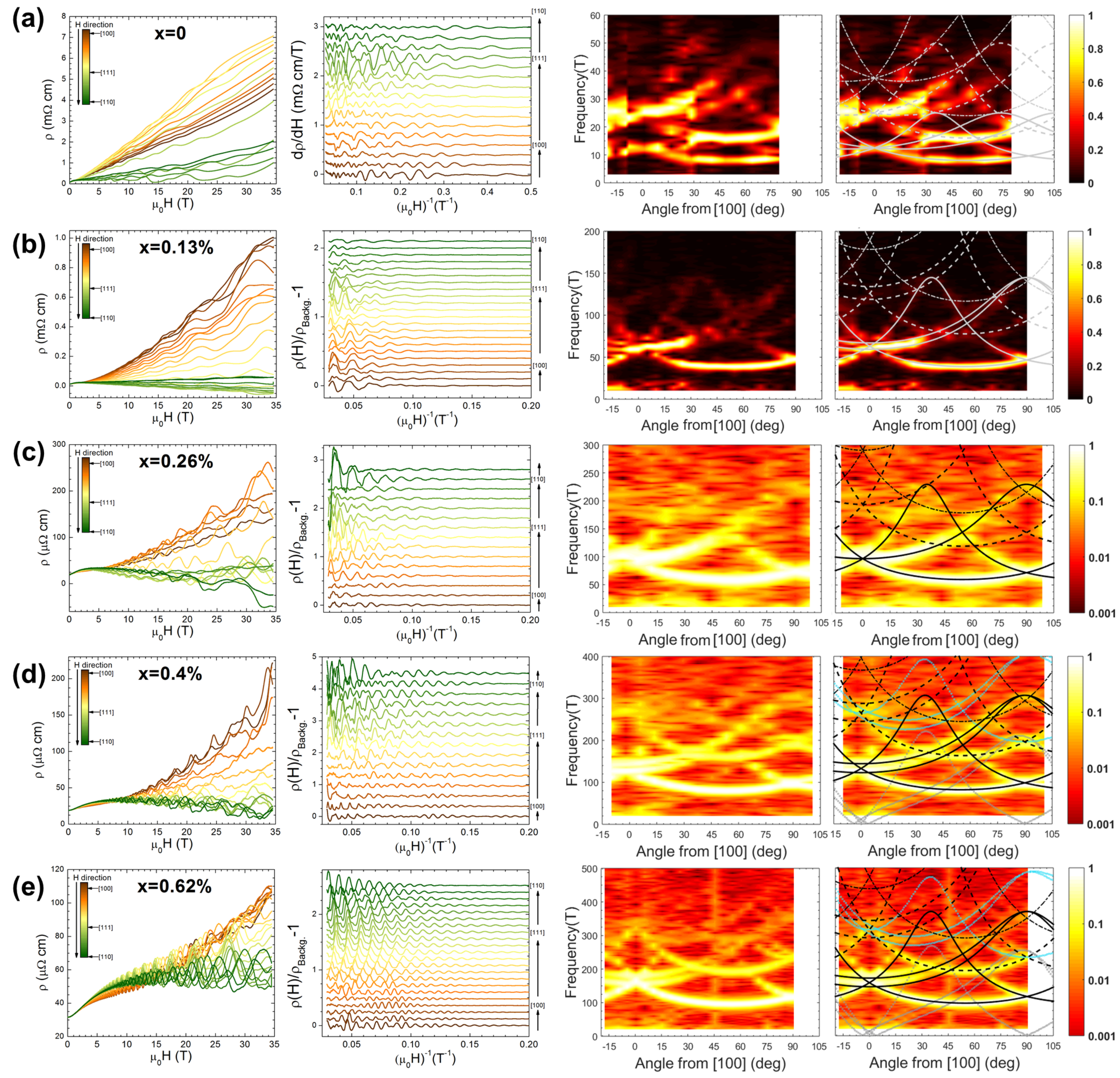} 
\vspace{0.2cm}
\caption{(Color online) Magnetoresistance measurements for Pb$_{1-x}$Na$_x$Te samples of different Na concentrations (row \textbf{(a)} $x=$0, row \textbf{(b)} $x=$0.13$\%$, row \textbf{(c)} $x=$0.26$\%$, row \textbf{(d)} $x=$0.4$\%$ and row \textbf{(e)} $x=$0.62$\%$) as a function of magnetic field, as rotated along the (110) plane. The first column shows the measured resistivity as a function of applied magnetic field. The second column shows the background-free resistivity, obtained as explained in the main text, as a function of inverse field. The third column shows the amplitude of the normalized 
FFT, represented by the color scale, as a function of the angle of the magnetic field from the [100] direction (horizontal axis), and the frequency (vertical axis). The last column 
replots column three, with a comparison to a perfect ellipsoidal model calculation superimposed (solid-lines for fundamental frequencies, and dashed-lines for higher-harmonics). The parameters used for the perfect ellipsoidal model calculation for each set of data are summarized in table \ref{table_Naparam}. For samples with $x=$0.13$\%$, 0.4$\%$ and 0.62$\%$, small deviations from the (110) plane of rotation are evidenced in the splitting of the angle evolution of the intermediate branch, and they were considered in the perfect ellipsoidal model comparison. For the two highest concentrations, combination frequency terms due to magnetic interaction effects are observed. These are identified in the fourth column plots by the light-blue dotted-lines (sum of fundamental branches) and gray dotted-lines (difference of fundamental branches).
}\label{fig_SdHall}
%\vspace{-0.4cm}
\end{figure*}

\begin{figure*}[!t]
%\vspace{-0.3cm}
\hspace{-0.3cm}
\centering
\includegraphics[scale=0.5]{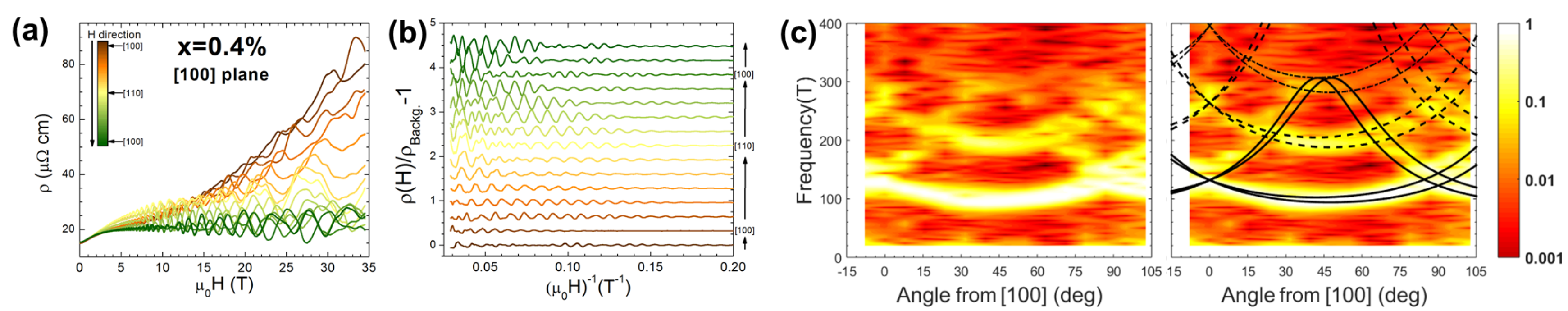} 
\caption{(Color online) {\bf (a)} Longitudinal magnetoresistance for a Na-doped PbTe sample with $x=$0.4$\%$ and Hall number $p_H=6.3\times 10^{19}$cm$^{-3}$, for different directions of the applied magnetic field, with respect to the [100] crystalline axis, as the field is rotated in the (100) plane. {\bf (b)} As in (a), as a function of inverse magnetic field, after eliminating the background, therefore only preserving the oscillatory part. {\bf (c)} The color scale in both plots represents the amplitude of the Fourier transform of the data shown in (b), as a function of the angle from the [100] direction (horizontal axis), and the frequency (vertical axis). For these plots, the field is rotated in the (100) plane. The right hand side figure replots the figure in the left, but with a perfect ellipsoidal model calculation superimposed on the data, up to the third harmonic (black lines). For the model, the plane of rotation is offset by 5.5$^{\circ}$ (about the [100] axis). The parameters used for the calculations are the same as those used for the (110) plane of rotation data in Fig. \ref{fig_SdHall}(d): $f_{min}=81.4$T and $f_{max}=307$T.}\label{fig_SdH4524100}
%\vspace{-0.4cm}
\end{figure*}

\begin{figure}[!htbp]
\centering
\includegraphics[scale=0.35]{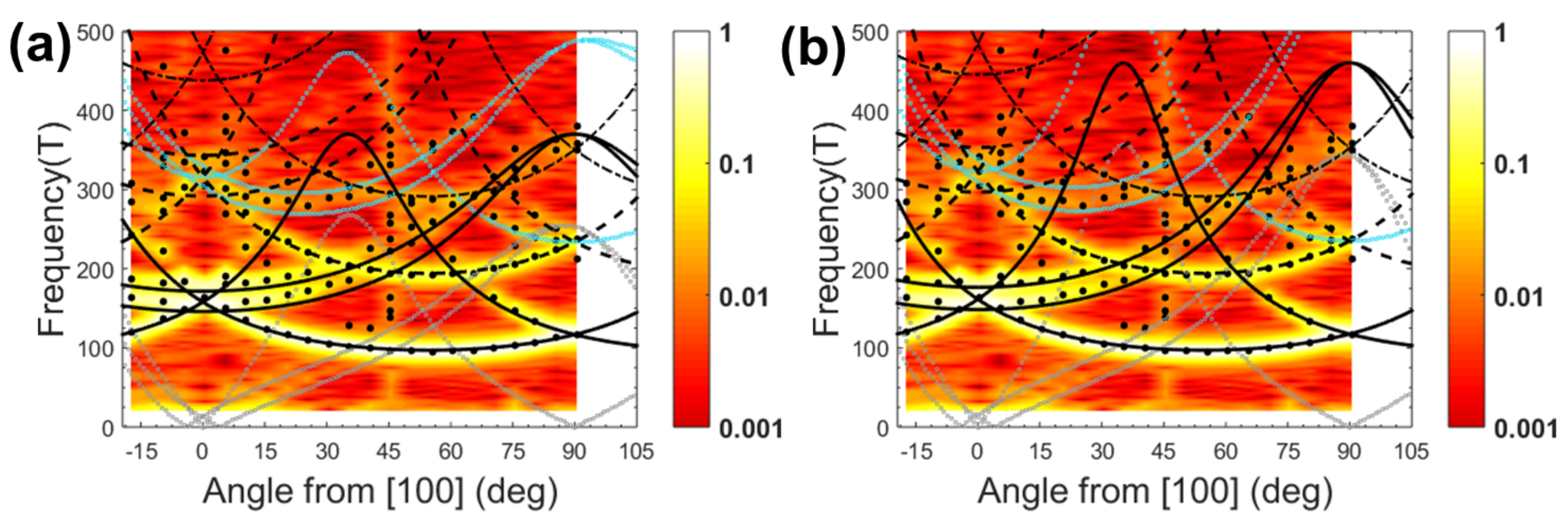}
%\vspace{-0.2cm}
\caption{(Color online) FFT of the background-free resistivity data of Fig. \ref{fig_SdHall}(e), as a function of the angle from the [100] direction and the frequency. A perfect ellipsoidal model calculation has been superimposed on the data, up to the third harmonic (black lines). In order to better guide the comparison with the perfect ellipsoidal model, the exact frequencies of the local maxima of the FFT for each angle (labeling only FFT peaks with amplitude 1$\%$ or more of the largest peak for each angle) are indicated by black-dots. The parameters used in the perfect ellipsoidal model for each plot are: {\bf (a)} $f_{min}=97$ T, $f_{max}=370$ T; and {\bf (b)} $f_{min}=97$ T, $f_{max}=460$ T. For both plots, an offset of 4$^{o}$ from the (110) plane of rotation (about the [110] axis) is considered, to account for the splitting seen in the middle branch. Additionally, the combination frequency terms are shown in light-blue-dotted lines (sum of fundamental branches) and gray-dotted lines (difference of fundamental branches). None of the fits presented here give a satisfactory description of the data, suggesting deviations from perfect ellipsoidicity.}\label{fig_SdH4504}
%\vspace{-0.4cm}
\end{figure}

\section{Experimental Results}

We divide the results section into two parts: In subsection A we show the angle dependence of the magnetoresistance as the magnetic field is rotated within a high symmetry crystallographic plane, and temperature is held fixed at $(1.5\pm 0.2)$ K. This allows us to obtain information about the topology of the Fermi surface and its evolution with carrier concentration. In subsection B we present measurements of the temperature dependence of the amplitude of oscillations in magnetoresistance along different high symmetry directions, in order to extract information about the effective cyclotron masses, and their evolution with carrier concentration.   

\subsection{High-field magnetoresistance measurements}

All data presented in this section were taken at a temperature of $(1.5\pm 0.2)$ K. For all the samples measured, large Shubnikov-de Haas (SdH) oscillations in magnetoresistance were observed starting at a field of approximately 4T for most samples. 
The first column of Figure \ref{fig_SdHall} shows symmetrized measurements of resistivity, $\rho$, as a function of magnetic field for Pb$_{1-x}$Na$_x$Te with (a) $x$=0 ($p_H=1.9\times 10^{18}$cm$^{-3}$), (b) $x$=0.13$\%$ ($p_H=2.1\times 10^{19}$cm$^{-3}$), (c) $x$=0.26$\%$ ($p_H=4.1\times 10^{19}$cm$^{-3}$), (d) $x$=0.4$\%$ ($p_H=6.3\times 10^{19}$cm$^{-3}$) and (e) $x$=0.62$\%$ ($p_H=9.4\times 10^{19}$cm$^{-3}$), for different field orientations in the (110) plane. As mentioned in section \ref{sec_DFTSdH}, the (110) plane is a natural plane to study the angle evolution of the SdH frequencies for this material, given that, in a perfect ellipsoidal scenario, it allows the determination of all the extremal cross-sectional areas of both, L- and $\Sigma$-pockets. The second column of Fig.~\ref{fig_SdHall} shows the oscillating component of the respective magnetoresistance curves, as a function of inverse field, extracted after the following background elimination procedure: for such low carrier densities, which imply low frequencies of oscillation, the determination of the frequencies and the tracking of their evolution with angle is challenging, given that only a few periods of oscillations are observed for the field range used, and additionally, several artifacts coming from background subtractions have characteristic frequencies that are comparable to the frequencies of interest. In our data analysis, several methods for background subtraction were tested. The method that generated the best resolution in the fast Fourier transform (FFT) for all the Na-doped samples, and that we use here, was a cubic-spline fitting of the non-oscillating component. For the self-doped $x$=0 sample, which is the sample with the lowest characteristic frequencies (as low as 8 T), the method that allowed the best resolution of the evolution of fundamental frequency branches was the computation of the first derivative.

The evolution with angle of the frequencies of oscillation is shown in the contour plots of the third and fourth columns of Fig. \ref{fig_SdHall}. The color scale for these plots represents the amplitude of the FFT of the corresponding curves in the second column, normalized by the maximum value of the FFT at each angle, as a function of the angle from the [100] direction, and frequency. For all samples, the fundamental frequency of the three expected branches of frequency evolution is clearly observed, and for some of the branches, the second and third harmonic can be identified. For the $x$=0 sample, the second harmonic seems to be stronger in amplitude than the fundamental, for all three branches. This effect is likely associated with the difficulty of resolving low frequency signals. For all samples, the branch that lies in the low frequency region for all angles contributes the dominant frequency in the magnetoresistance, which is associated with its higher mobility with respect to the other two branches. For the higher concentration samples, the high frequency contributions are weaker, and a logarithmic scale in the contour plots is used in order to highlight their angle evolution. In order to determine the characteristic frequencies of oscillation, and the possible deviations of the Fermi surface from a perfect ellipsoidal model, a comparison of these plots with the frequency evolution for a Fermi surface containing eight half-ellipsoids at the L point (perfect ellipsoidal model) is shown in the fourth column plots of Fig. \ref{fig_SdHall}. The fundamental frequencies, as well as the second and third harmonics are shown for each sample. The splitting seen in the intermediate frequency branch for most of the samples can be successfully accounted for by a small offset in the plane of rotation. For the $x$=0 sample, an offset of 12$^{\circ}$ about the [001] axis was considered in the perfect ellipsoidal model. For samples with $x=$0.13$\%$, the offset is 3$^{\circ}$ about the [110] axis; and for $x=$0.4$\%$ and 0.62$\%$, the offset is 4$^{\circ}$ about the [110] axis. 

\begin{table*}[!htbp]
\centering
%\begin{small}
\caption{Fermi surface parameters for Na-doped PbTe, obtained from comparison between our measured data and a perfect ellipsoidal model.}\label{table_Naparam}
\begin{ruledtabular}
\begin{tabular}{ c c c c c c c }
	$x$(at.$\%$) & $p_H$($\times 10^{19}$cm$^{-3}$) & $f_{min}$ (T) & $f_{[100]}$ (T) & $f_{max}$ (T) & $K$ & $p_{FS-Vol}$ ($\times 10^{19}$cm$^{-3}$)   \tabularnewline\hline
	0 & 0.19 $\pm$ 0.001 & 8 $\pm$ 1 & 12.5 $\pm$ 2 & 25 $\pm$ 2 & 10 $\pm$ 4 & 0.16 $\pm$ 0.02 \tabularnewline 
	0.04 & 0.75 $\pm$ 0.01 & 17 $\pm$ 5 & 34 $\pm$ 7 & - & - & - \\ 
	0.13 & 2.09 $\pm$ 0.01 & 39 $\pm$ 4 & 63 $\pm$ 5 & 145 $\pm$ 7 & 14 $\pm$ 3 & 2.1 $\pm$ 0.2 \tabularnewline
	0.26 & 4.1 $\pm$ 0.06 & 60 $\pm$ 8 & 97 $\pm$ 10 & 230 $\pm$ 7 & 15 $\pm$ 4 & 4.0 $\pm$ 0.3 \tabularnewline 
	0.4 & 6.3 $\pm$ 0.6 & 81 $\pm$ 4 & 132 $\pm$ 13 & 307 $\pm$ 6 & 14 $\pm$ 2 & 6.3 $\pm$ 0.2 \tabularnewline 
	0.62 & 9.4 $\pm$ 0.6 & 97 $\pm$ 12 & 157.5 $\pm$ 16 & 370 $\pm$ 90 & 15 $\pm$ 8 & 8.3 $\pm$ 2.1 \tabularnewline
\end{tabular}
\end{ruledtabular}
%\end{small} 

\end{table*}

The parameters of minimum and maximum cross-sectional areas ($f_{min}$ and $f_{max}$) used in the perfect ellipsoidal model comparison for each sample are summarized in Table \ref{table_Naparam}. The minimum cross-sectional area of the L-pockets, associated with $f_{min}$, can be determined very accurately from the value of the fundamental frequency of oscillation at 55$^{\circ}$ from the [100] direction in the (110) plane, which is clearly observed for all the samples measured. Additionally, the maximum cross sectional area of the L-pockets, associated with $f_{max}$, can be directly observed in the FFT plots of samples with Na concentration up to 0.4$\%$. Also, up to this concentration, the matching between the angle evolution of the frequencies of oscillation with that expected for a perfect ellipsoidal model is satisfactory. Nevertheless, for this last concentration, the maximum frequency of the ellipsoids is resolvable close to 90$^{\circ}$ from [100], but becomes blurred close to 35$^{\circ}$. Therefore, although the value of the maximum frequency can be determined from the 90$^{\circ}$ area, possible deviations from ellipsoidal model that could be identified around 35$^{\circ}$ cannot be resolved. However, given the round shape of the upper-branch around 90$^{\circ}$, we can say that features associated with possible departures from the  ellipsoidal model are not observed (see Fig. \ref{fig_boris}(c)). This last statement is confirmed by magnetoresistance measurements in an additional sample of the same batch as the field is rotated along the (100) plane, as shown in Fig. \ref{fig_SdH4524100}. The comparison of the FFT angle evolution and the perfect ellipsoidal model, using the same extremal cross-sectional area parameters as for the measurements with field along the (110) plane, confirms the matching of the data with the perfect ellipsoidal model for samples of this Na composition ($x$=0.4$\%$). For the highest Na concentration sample measured, $x$=0.62$\%$, possible deviations from perfect ellipsoidicity are observed, and will be discussed later in this section. 

As can be seen in the third and fourth columns of Figs. \ref{fig_SdHall}(d) and (e), additional features in the angle dependence plots occur for the two highest Na-doped samples. Nevertheless, all of these features can be identified as the sum and difference of the fundamental frequencies of the L-pockets, as can be observed in the light-blue and gray curves in the fourth column plots of Figs. \ref{fig_SdHall}(d) and \ref{fig_SdHall}(e). The presence of such combination frequencies can be attributed to magnetic interaction (MI) effects, expected when the amplitude of the oscillating component of the magnetization, $\tilde{M}$ is comparable to $H^2/f$, in such a way that the total magnetic field $\vec{B}=\vec{H} + 4\pi\vec{M}$ and not just $\vec{H}$, needs to be considered in the Lifshitz-Kosevich (LK) formalism of quantum oscillations \cite{Shoenberg} (see appendix \ref{app_QO}).

As was suggested above, the sample with the highest Na concentration studied in this work, $x=$0.62$\%$, shows possible indications of deviations from perfect ellipsoidicity. For this sample the high frequency components of the oscillations are blurred, and the evolution of the different branches can be observed only up to 400 T. As we mentioned previously, the determination of $f_{min}$ for all samples has a very low uncertainty, particularly for this sample, given that we can clearly observe up to the third-harmonic of the lower branch (see fig. \ref{fig_SdHall}(e)). Fixing this value to $f_{min}=97$ T, Figure \ref{fig_SdH4504} shows a comparison between the angle evolution of the frequencies of oscillation for this sample, and a perfect ellipsoidal model using two different values of $f_{max}$. In order to guide the comparison better, both plots in this figure show the exact frequency positions of the maxima of the FFT peaks for all angles (in black-filled circles). Around the angle of 90$^{\circ}$ we observe some weight in the FFT (yellow color) around 350-370 T, which we could interpret as an indication of the value of $f_{max}$. This value is the one used in the perfect ellipsoidal model in Fig. \ref{fig_SdH4504}(a) (as well as Fig. \ref{fig_SdHall}(e)). In this figure, we can see that the matching between the data and the perfect ellipsoidal model is not satisfactory, especially close to the 0$^{\circ}$ area of the plot. Interestingly, the 90$^{\circ}$-370 T area overlaps with the region at which the third harmonic of the lower branch passes. This could indicate that the weight observed at this region belongs to this third harmonic, and not to $f_{max}$. Figure \ref{fig_SdH4504}(b) shows a comparison between the data and a perfect ellipsoidal model using the same $f_{min}=97$ T, but now using a larger value of $f_{max}=460$ T. These values provide a better matching between the data and a perfect ellipsoidal model for the region of 0$^{\circ}$. However, the combination frequency terms, due to magnetic-interaction effects, suggest that this fit is not satisfactory, as the evolution of the combination frequency data points around 60$^{\circ}$-350 T seems to be less steep, being better matched by the fit using $f_{max}=370$ T, as shown in Fig. \ref{fig_SdH4504}(a). The lack of a satisfactory perfect ellipsoidal model to describe the data can be interpreted as deviations from perfect ellipsoidicity of the L-pockets for this Na concentration. The mismatch of the data and the ellipsoidal model is observed in the intermediate branch, which is consistent with the guidelines given by the DFT calculations.

For all the samples measured, the only features observed in the angle evolution of the frequencies of oscillations come from the L-pockets. Furthermore, the carrier concentration calculated from Luttinger's theorem and the volume in $k$-space of the L-pockets, obtained through the comparison of the FFT evolution and the perfect ellipsoidal model, which we label as $p_{FS-Vol}$, matches perfectly (within the error bars) with the Hall number (equivalent to the carrier concentration for a single band compound) for all Na-doped samples up to $x=$0.4$\%$, as shown in table \ref{table_Naparam}. This fact confirms that the only band contributing to conduction in this compound up to this Na-concentration is the L band. Moreover, the small mismatch between the L-pocket Luttinger volume and the Hall number for the highest Na concentration sample, $x=$0.62$\%$ presumably comes from deviations from perfect ellipsoidicity, as discussed.

\begin{figure}[!t]
%%\vspace{-0.3cm}
%%\hspace{-0.2cm}
\centering
\includegraphics[scale=0.6]{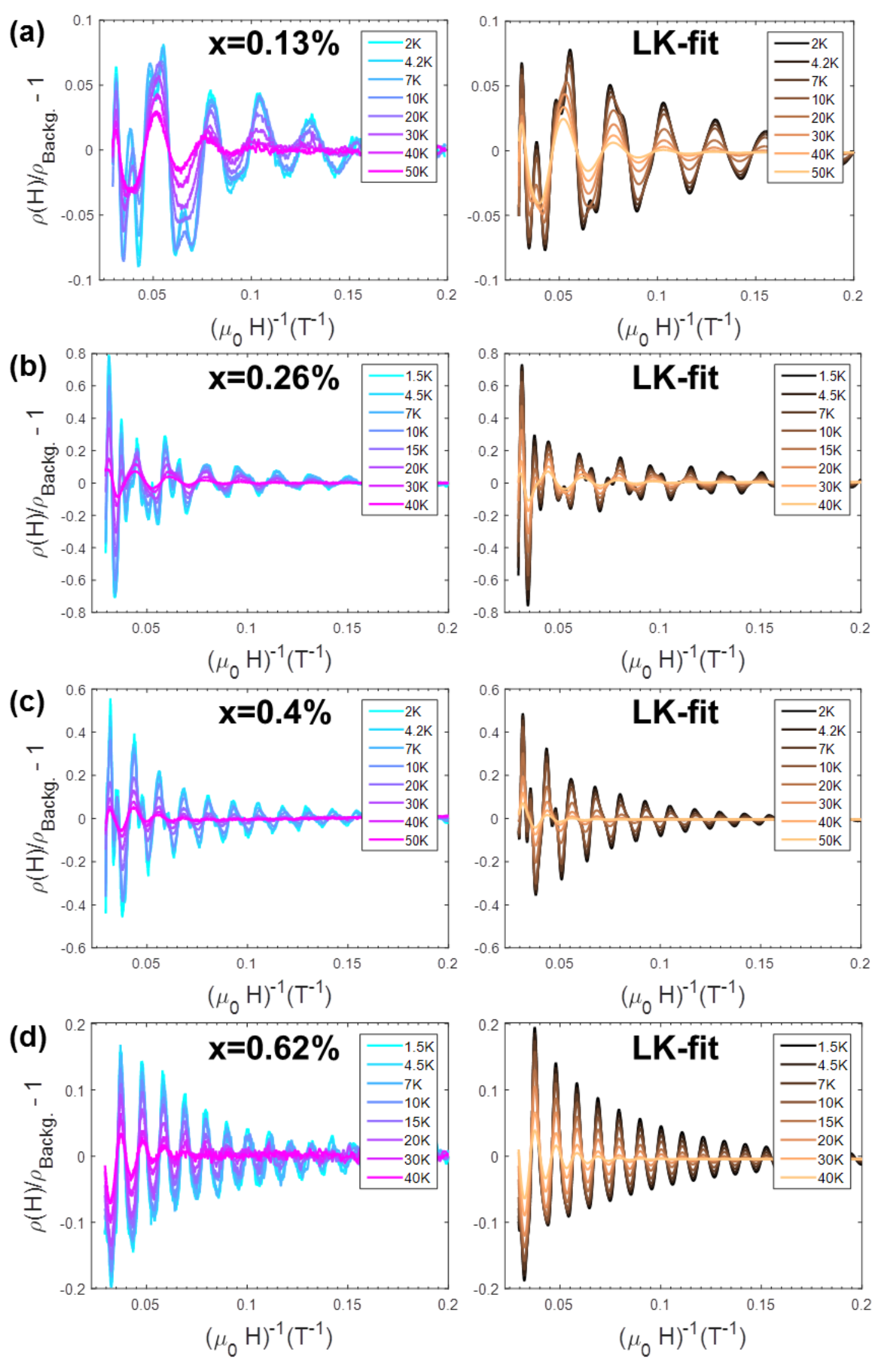}
\caption{(Color online) Temperature dependence of the amplitude of the oscillating component of magnetoresistance for Pb$_{1-x}$Na$_x$Te samples, with magnetic field along the [111] direction (55$^{\circ}$ from the [100] direction, in the (110) plane). The left-column plots show the background-free data at different temperatures. The right-column plots show the fits of the data to the LK-formula in equation \ref{eq_LKmass}, using the four most dominant frequencies observed in the FFT of the lowest temperature curve. From this fit, the values of cyclotron effective mass and Dingle temperature, for each frequency term, are obtained.
}\label{fig_Tdepall111}
%\vspace{-0.4cm}
\end{figure}

\begin{figure}[!t]
\centering
%%\hspace{-0.6cm}
\includegraphics[scale=0.61]{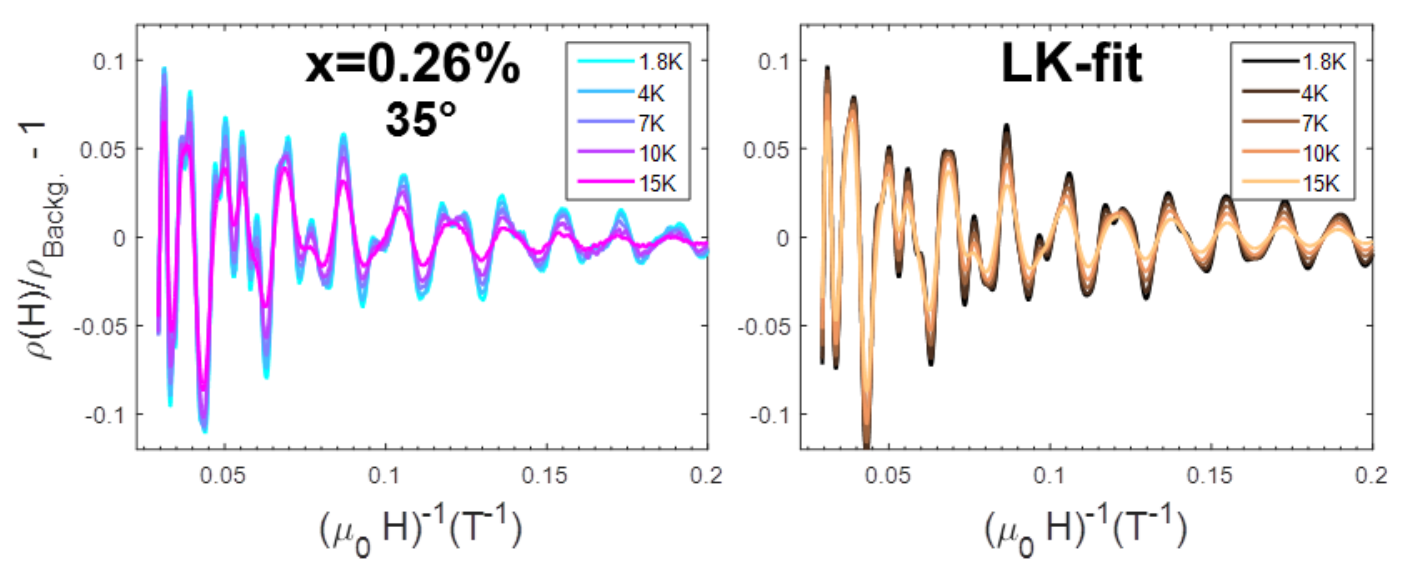} %\hspace{-0.7cm}
\caption{(Color online) Temperature dependence of the amplitude of the oscillating component of magnetoresistance for a Pb$_{1-x}$Na$_x$Te sample with $x=$0.24$\%$, and magnetic field oriented close to 35$^{\circ}$ from the [100] direction, along the (110) plane. For this orientation, the cross-sectional area of two of eight L-pockets corresponds to the maximum cross-sectional area of the ellipsoids. The left-column plot shows the background-free data at different temperatures. The right-column plot shows the fit of the data to the LK-formula in equation \ref{eq_LKmass}, using the five most dominant frequencies observed in the FFT of the lowest temperature curve. From this fit, the values of cyclotron effective mass and Dingle temperature, for each frequency term, are obtained.}\label{fig_Tdepall35}
\end{figure}

\begin{table*}[!t]
\centering
%\begin{small}
\caption{Cyclotron effective masses for Pb$_{1-x}$Na$_x$Te samples along different high symmetry directions. These parameters were obtained through fitting of the curves in Figs. \ref{fig_Tdepall111}, \ref{fig_Tdepall35} and \ref{fig_Tdepall100} to the LK-formula in equation \ref{eq_LKmass}.}\label{table_mass}
\begin{ruledtabular}
\begin{tabular}{ccccccc}
	$x$(at.$\%$) & p$_H$($\times 10^{19}$cm$^{-3}$) & $\Theta_{D,\bot}$ (K) & $m^{cyc}_{\bot}/m_e$  & $m^{cyc}_{[100]}/m_e$ & $m^{cyc}_{\|}/m_e$ \tabularnewline\hline
	0.04 & 0.75 $\pm$ 0.007 & - & - & 0.098 $\pm$ 0.001 & - \tabularnewline
	0.13 & 2.09 $\pm$ 0.006 & 9 $\pm$ 4 & 0.068 $\pm$ 0.007 & 0.085 $\pm$ 0.001 & - \tabularnewline
	0.26 & 4.1 $\pm$ 0.06 & 10 $\pm$ 3 & 0.089 $\pm$ 0.002 & 0.15 $\pm$ 0.01 & 0.29 $\pm$ 0.04 \tabularnewline
	0.4 & 6.3 $\pm$ 0.6 & 9.9 $\pm$ 0.2 & 0.14 $\pm$ 0.03 & 0.172 $\pm$ 0.004 & - \tabularnewline
	0.62 & 9.4 $\pm$ 0.6 & 9.5 $\pm$ 0.8 & 0.13 $\pm$ 0.02 & 0.225 $\pm$ 0.006 & - \tabularnewline
\end{tabular}
\end{ruledtabular}
%\end{small} 
\end{table*}

\subsection{Temperature dependence of Quantum Oscillations} \label{sec_Tdep}

In order to determine the effective cyclotron mass of holes in Na-doped PbTe, and their evolution with carrier concentration, the temperature dependence of the oscillation amplitude was measured for samples of different Na concentrations, with the field oriented along or close to high symmetry crystallographic directions. The cyclotron effective masses were obtained by simultaneous fitting of the curves for all temperatures to the Lifshitz-Kosevich (LK) formula (in SI units) \cite{Shoenberg}

\begin{eqnarray}
\frac{\rho(H) -\rho_{0}}{\rho_0} & & =\sum_{i} C_i \left\{\exp\left(\frac{-\num{14.7}(m_i^{cyc}/m_e) \Theta_{D,i})}{H}\right)\right\} \nonumber \\
& & \times\left\{\frac{T/H}{\sinh\left(\num{14.7}(m_i^{cyc}/m_e)T/H\right)}\right\} \nonumber \\
& & \times\cos\left[2\pi\frac{f_i}{H}+\phi_i\right] \label{eq_LKmass}
\end{eqnarray}

where the sum is over the frequencies observed in the data, and for which a separate cyclotron effective mass, $m_i^{cyc}/m_e$ and Dingle temperature, $\Theta_{D,i}$ can be obtained for each frequency term. This method of extracting the cyclotron effective mass, through direct fitting to the LK formula, is required for an accurate determination of these quantities for such a low carrier density material. For low frequency oscillations, the number of periods observed in the given field range is limited, resulting in FFTs with amplitudes highly dependent on windowing effects, variations in field range or variations in signal sampling. In contrast to the fitting of the FFT amplitudes to the LK formula, the method widely used for the determination of effective masses of higher carrier concentration metals, the values of effective masses obtained though a direct fitting of the data to the LK formula are robust to such variations. 

Figure \ref{fig_Tdepall111} shows the temperature dependence of the oscillating component of magnetoresistance for Pb$_{1-x}$Na$_x$Te samples of different Na concentrations, for field oriented along the [111] direction, which provides direct access to the transverse cyclotron effective mass, $m_{\bot}^{cyc}$, associated with the minimum cross-sectional area of the L-pockets. Least-squares fits to equation \ref{eq_LKmass}, including up to the fourth strongest frequency component, for each Na doping, and for a field range of 5 T to 34 T, are shown in the right-column plots of this figure. The cyclotron masses and Dingle temperatures obtained for the fundamental frequency, i.e., $m_{\bot}^{cyc}$ and $\Theta_{D,\bot}$, as a function of carrier concentration, are summarized in Table \ref{table_mass}, and plotted in Fig. \ref{fig_massAll} and Fig. \ref{fig_TDingle}, in the discussion section.

Additionally, Fig. \ref{fig_Tdepall35} shows magnetoresistance curves at different temperatures for a sample with Na concentration of 0.26$\%$, with the magnetic field oriented close to 35$^\circ$ from the [100] direction, in the (110) plane. For such field orientation, one of the Fermi surface cross-sectional areas corresponds to the maximum cross sectional area of the ellipsoids (in a perfect ellipsoidal model), associated with the maximum or longitudinal cyclotron mass, $m_{||}^{cyc}$. From this measurements, cyclotron masses along intermediate directions can also be found, and these are presented in Fig. \ref{fig_mangle4505} of the discussion section.

\section{Discussion}

\subsection{Fermi surface topology} \label{sec_topo}

\begin{figure}[!htbp]
\vspace{-0.3cm}
\centering
\includegraphics[scale=0.32]{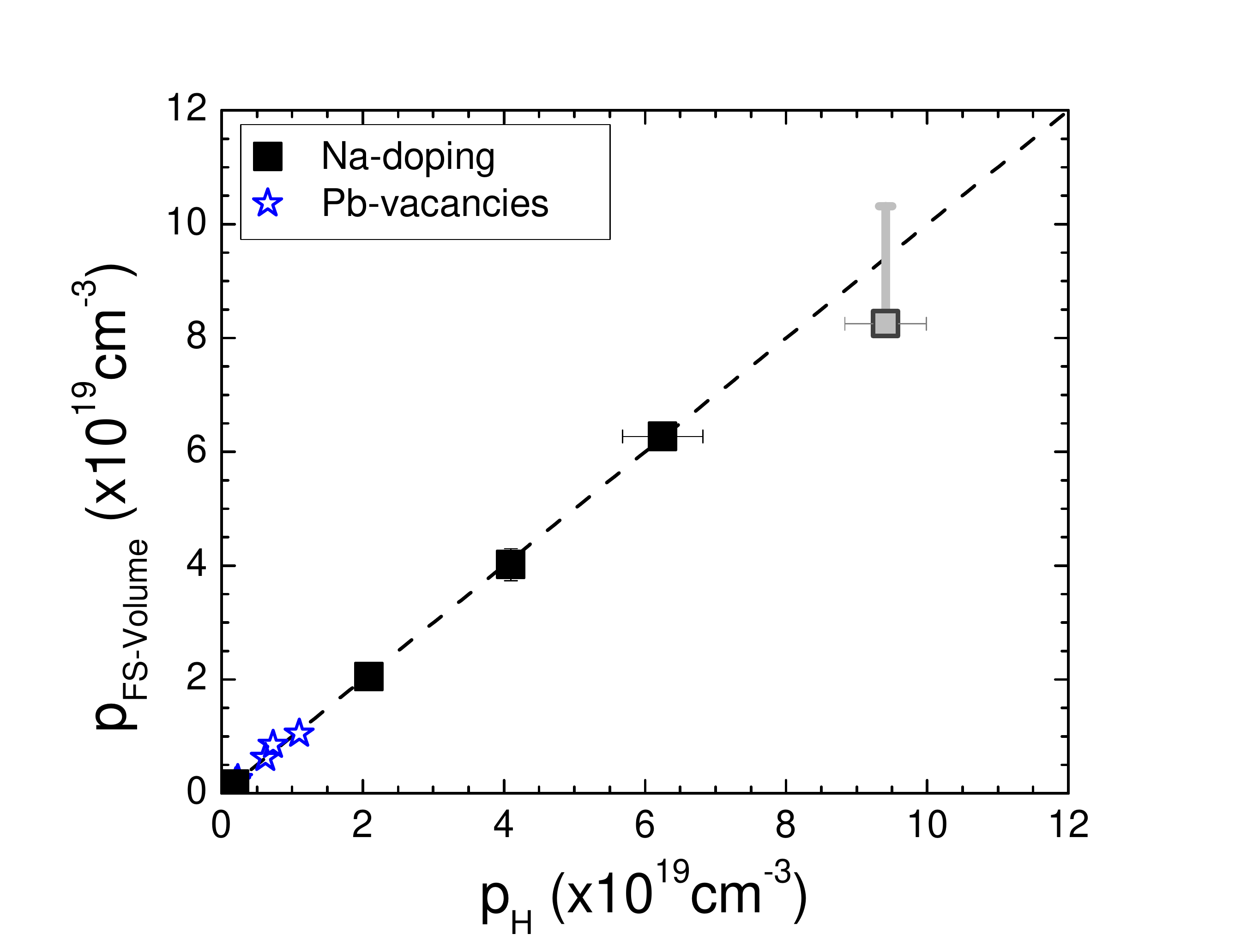} 
\caption{(Color online) Carrier concentration calculated from Luttinger's theorem and the volume of the L-pockets extracted from the comparison between the data and a perfect ellipsoidal model, as a function of the Hall number, for Na-doped PbTe (black squares), and obtained using the ellipsoid parameters from previous studies in refs. \citenum{burke2} and \citenum{burke1} (blue stars). The dashed line shows the expected behavior for a single-parabolic band, for which the carrier density enclosed by the Fermi surface, as determined through Luttinger's theorem matches the carrier density measured using the Hall effect. All the measured samples lie on this line, and the deviations seen for the highest Na doping are attributed to deviations from perfect ellipsoidicity.}\label{fig_pHFSVolNa}
\end{figure}

\begin{figure*}[!htbp]
\centering
%\hspace{-0.6cm}
\centering
\minipage{0.333\textwidth}
\includegraphics[scale=0.25]{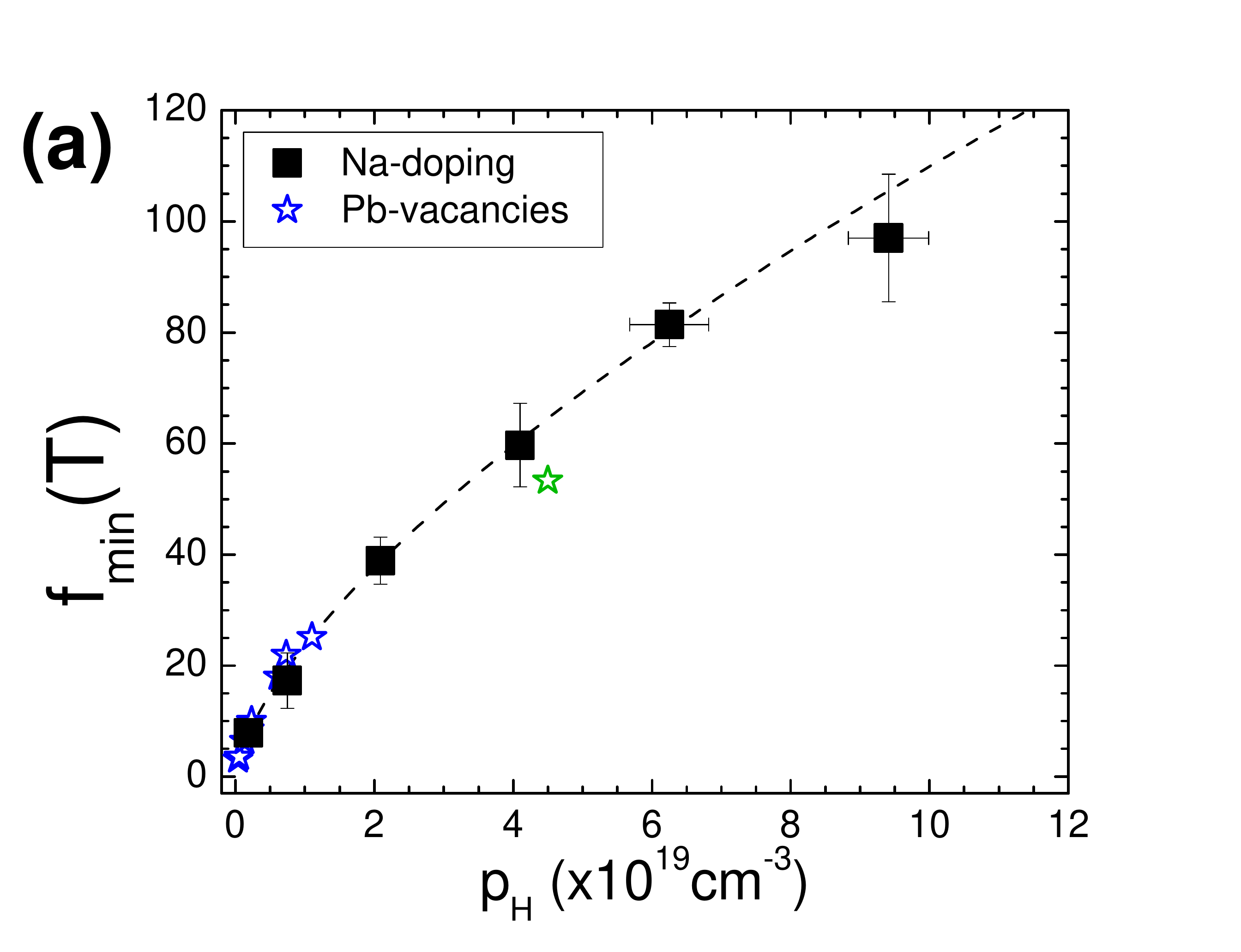} %\hspace{-0.7cm}
\endminipage\hfill
\centering
\minipage{0.333\textwidth}
%\hspace{-0.8cm}
\includegraphics[scale=0.25]{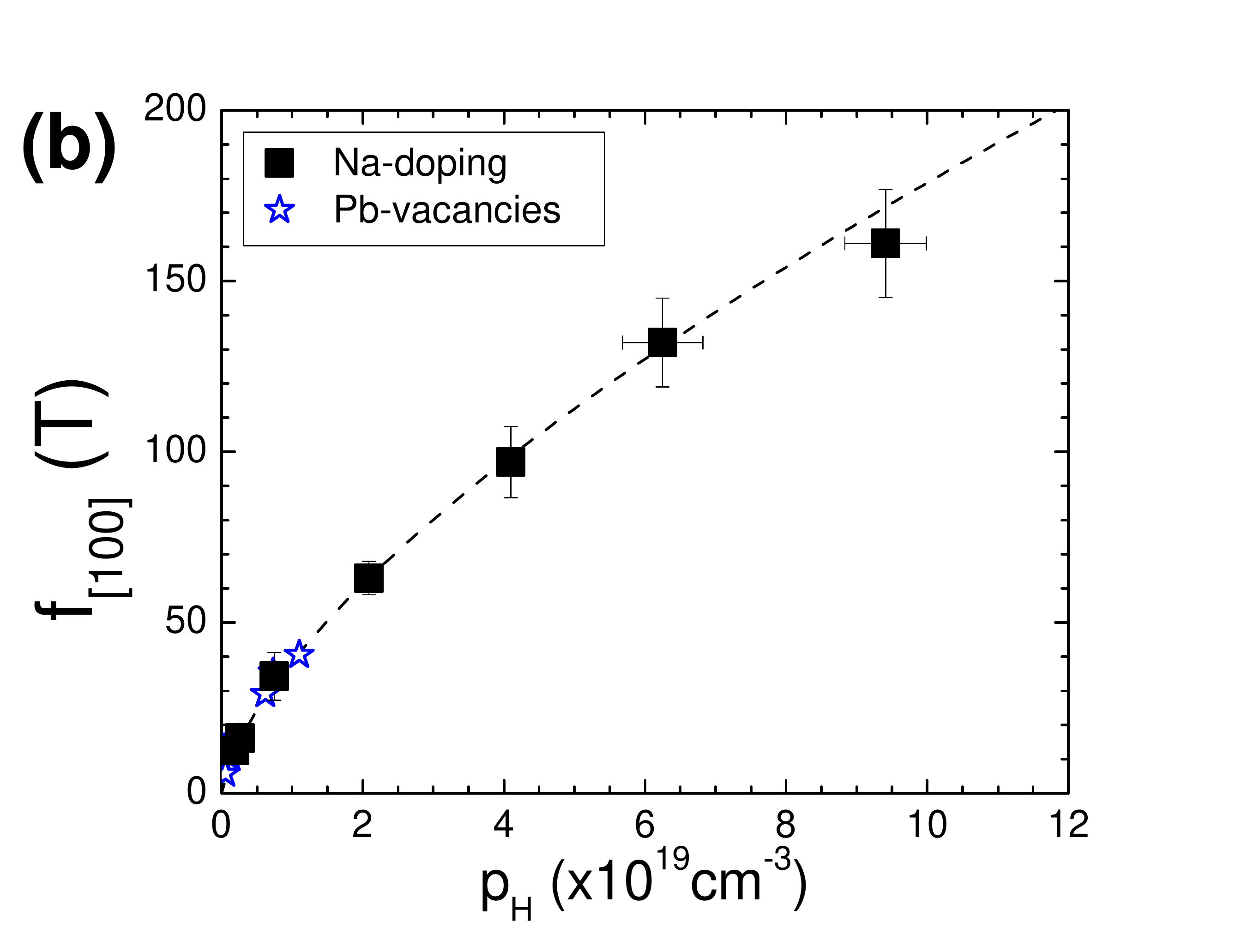} %\hspace{-0.7cm}
\endminipage\hfill
\centering
\minipage{0.333\textwidth}
%\hspace{-0.8cm}
\includegraphics[scale=0.25]{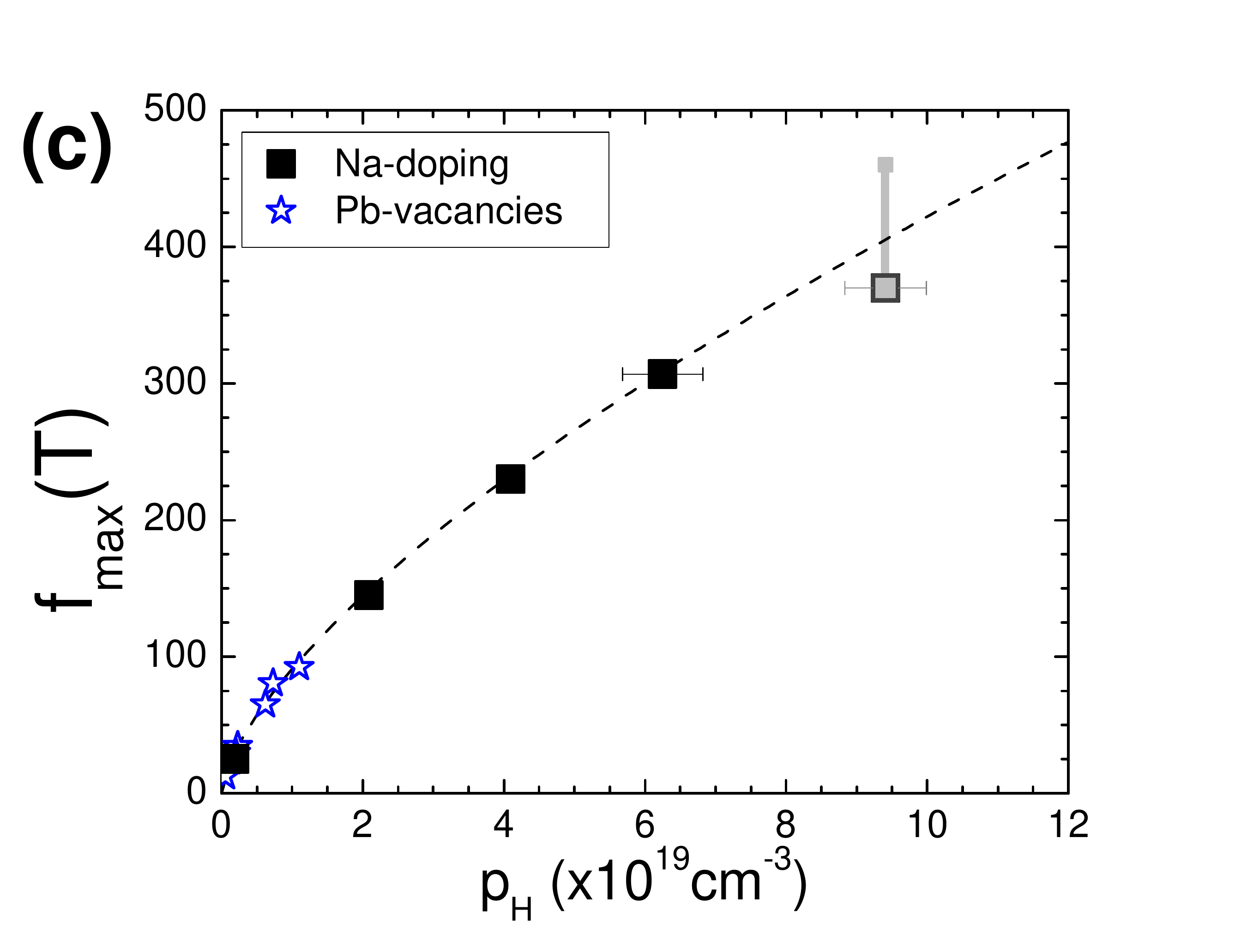} %\hspace{-0.7cm}
\endminipage\hfill
\caption{(Color online) Evolution of the characteristic frequencies of the L-pockets with Hall number, for Na doping, as determined from this study, and for self-doped samples from the works in refs. \citenum{burke2} and \citenum{burke1}: {\bf (a)} Frequency associated with the L-pockets' minimum cross-sectional area, $f_{min}$, {\bf (b)} Frequency associated with the L-pockets' cross-sectional area in the [100] direction, $f_{[100]}$, and {\bf (c)} Frequency associated with the L-pockets' maximum cross-sectional area, $f_{max}$. The blue-star symbols are data points obtained by previous quantum oscillation studies from other authors \cite{burke1,burke2}, in self-doped PbTe with different levels of Pb vacancies (the last star in $f_{min}$, in green, was obtained by Na doping). The dashed line in all the plots is the functional dependence of $p^{2/3}$ expected for a perfect ellipsoidal model with fixed anisotropy.}\label{fig_freqsNa}
\end{figure*}

\begin{figure}[!htbp]
\centering
%\hspace{-1cm}
\includegraphics[scale=0.3]{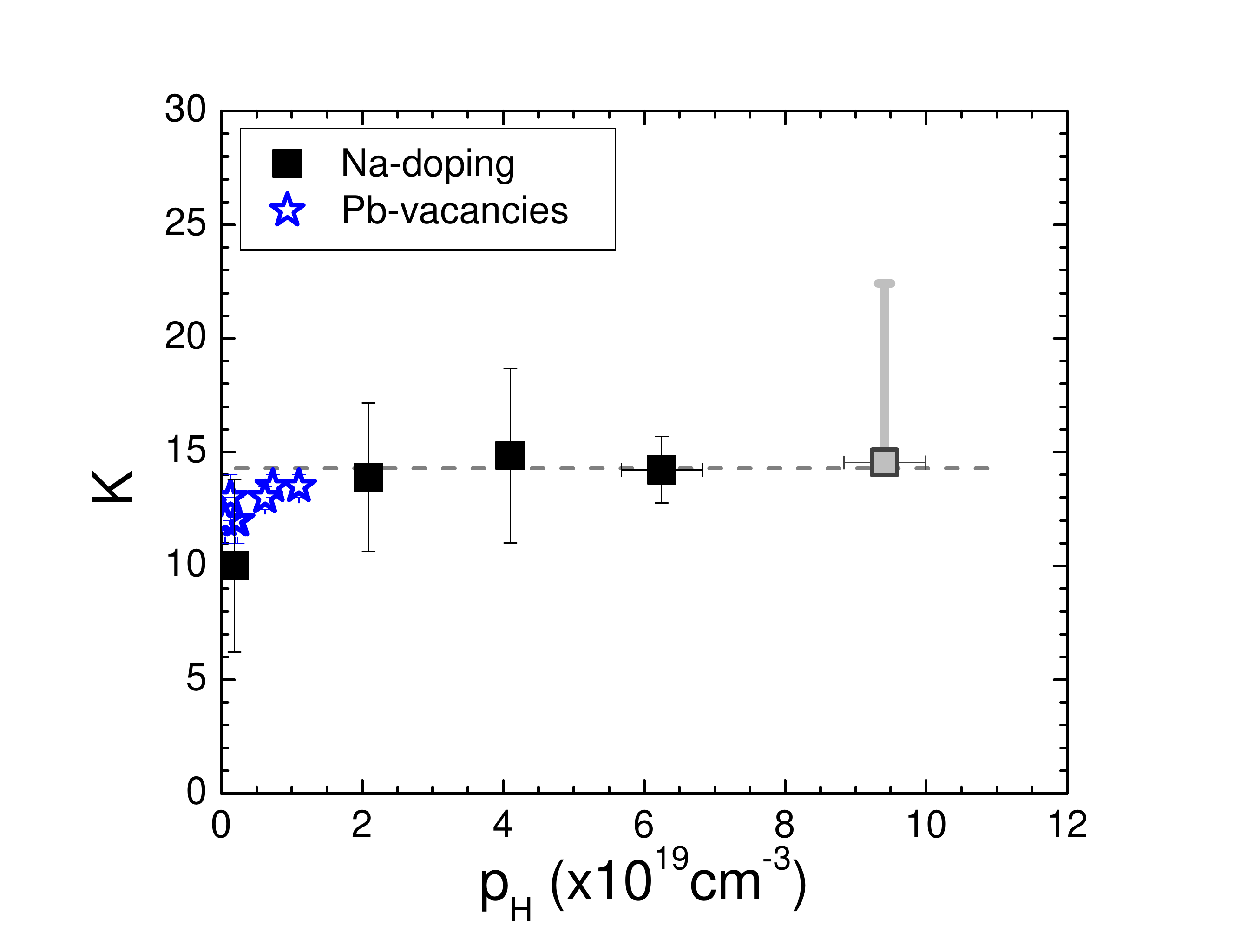} 
\caption[Anisotropy parameter of the L-pockets extracted from data, vs. Hall number]{Anisotropy parameter of the L-pockets, $K=(f_{max}/f_{min})^2$, extracted from the data, as a function of the Hall number for Na-doped samples, as determined from this study, and for self-doped samples from the works in refs. \citenum{burke2} and \citenum{burke1}. The horizontal gray-line shows the average value of $K=14.3\pm 0.4$ for this range of concentrations.}\label{fig_KNa}
\end{figure}

Having presented the data and the analysis performed to obtain the various Fermi surface parameters for different Na-doping levels, we now summarize them and present their evolution as a function of depth in the valence band. The parameters obtained in the previous section are summarized in table \ref{table_Naparam}, where we also include data from an additional Na composition ($x=0.04\%$) for which measurements in a more limited field range (up to 14 T) were taken. %, particularly, temperature dependence of the amplitude of oscillations, as it will be presented in the next section. 

Figure \ref{fig_pHFSVolNa} shows the L-pockets' Luttinger volume as a function of Hall number for the Na-doped PbTe samples studied, plus self-doped (by Pb vacancies) samples measured in previous SdH studies by other groups \cite{burke1,burke2}. For a single-parabolic-band model, these two quantities are expected to exactly match with each other, and to lie on the dashed line shown in the figure. This is indeed the case for all the samples studied, including the self-doped ones. The deviations seen for the last Na doping can be attributed to deviations from perfect ellipsoidicity of the pockets, as discussed above. The matching between the L-pockets' Luttinger volumes and Hall numbers implies that PbTe, up to a carrier concentration of $p_H=(9.4\pm 0.6)\times 10^{19}$cm$^{-3}$, is {\bf single band}, that is, all the carriers contributing to conduction belong to the L band. This result implies that the band offset between the L and $\Sigma$ valence band maxima is underestimated in our DFT calculations, as well as all previously published band-structure calculations \cite{Martinez1,Parker,Wiend1, Cho1,Salameh,Nemov1,Ravich, Ravich2,Ravich3}, which predict the appearance of the $\Sigma$ band at a hole concentration of the order of $p\approx1\times 10^{19}$cm$^{-3}$.

The evolution of the three high symmetry L-pocket cross-sectional areas, in frequency units ($f_{min}$, $f_{max}$ and $f_{[100]}$), with Hall number is plotted in figure \ref{fig_freqsNa}. For a perfect ellipsoidal model, all the cross-sectional areas are expected to scale with carrier concentration as $p_H^{2/3}$. This is in fact the functional form followed by most cross-sectional areas in fig. \ref{fig_freqsNa}, as shown by the dashed line. The last Na-doped sample deviates from this line, confirming the departure from perfect ellipsoidicity of the pockets for this high carrier concentration. However, for carrier concentrations below $p_H=6.3\times 10^{19}$cm$^{-3}$, we can say that the L-pockets are well described by a perfect ellipsoidal model, within the experimental resolution. For the highest Na concentration studied, the deviation from the perfect ellipsoidal behavior follows the expected trend predicted by our DFT calculations, as presented in Fig. \ref{fig_freq_nl_dft}. 

Additionally, the anisotropy of the L-pockets, $K=(f_{max}/f_{min})^2$, is approximately constant with carrier concentration ($K=14.3\pm 0.4$), for the range of carrier concentrations of interest, as shown in Fig. \ref{fig_KNa}. The observation of a constant anisotropy of the L-pockets with carrier concentration confirms previous results by Burke et al. \cite{burke1} for p-type self doped PbTe with carrier concentrations below $1\times10^{19}$cm$^{-3}$ (shown as blue stars in Fig. \ref{fig_KNa}), and contrasts the results by Cuff et al. \cite{Cuff1} in self-doped samples with carrier concentrations up to $~3\times10^{18}$cm$^{-3}$, in which a decrease in $K$ with increasing carrier concentration is observed. The $K$ values reported by Burke et al. are slightly less than the average value of 14.3$\pm$0.4 found in this work. However as discussed previously, an accurate estimation of the Fermi surface parameters for the low carrier concentration regime is challenging given the few periods of oscillation observed in a limited field range. This could be the reason for the lower $K$ value obtained for the $x=0$ sample measured in this work. A constant value of $K$ with carrier concentration is expected in a perfect parabolic band model, in which the L-pocket anisotropy is equivalent to the band mass anisotropy, $K=m_{\|}/m_{\bot}$, where $m_{\|}$ is the effective band mass along the ellipsoidal L-pocket major semi-axis (longitudinal band mass), and $m_{\bot}$ is the effective band mass along the ellipsoidal L-pocket minor semi-axis (transverse band mass) (in terms of the cyclotron effective masses, $K=(m^{cyc}_{\|}/m^{cyc}_{\bot})^2$, as shown in appendix \ref{app_massan}). However, a constant $K$ value can also be obtained for specific models with dispersion relations in which corrections for non-parabolicity of the band are considered, as we will present in the next section.

\subsection{Effective cyclotron masses and relaxation time}

\begin{figure}[!t]
\centering
%\hspace{-0.6cm}
\centering
\includegraphics[scale=0.3]{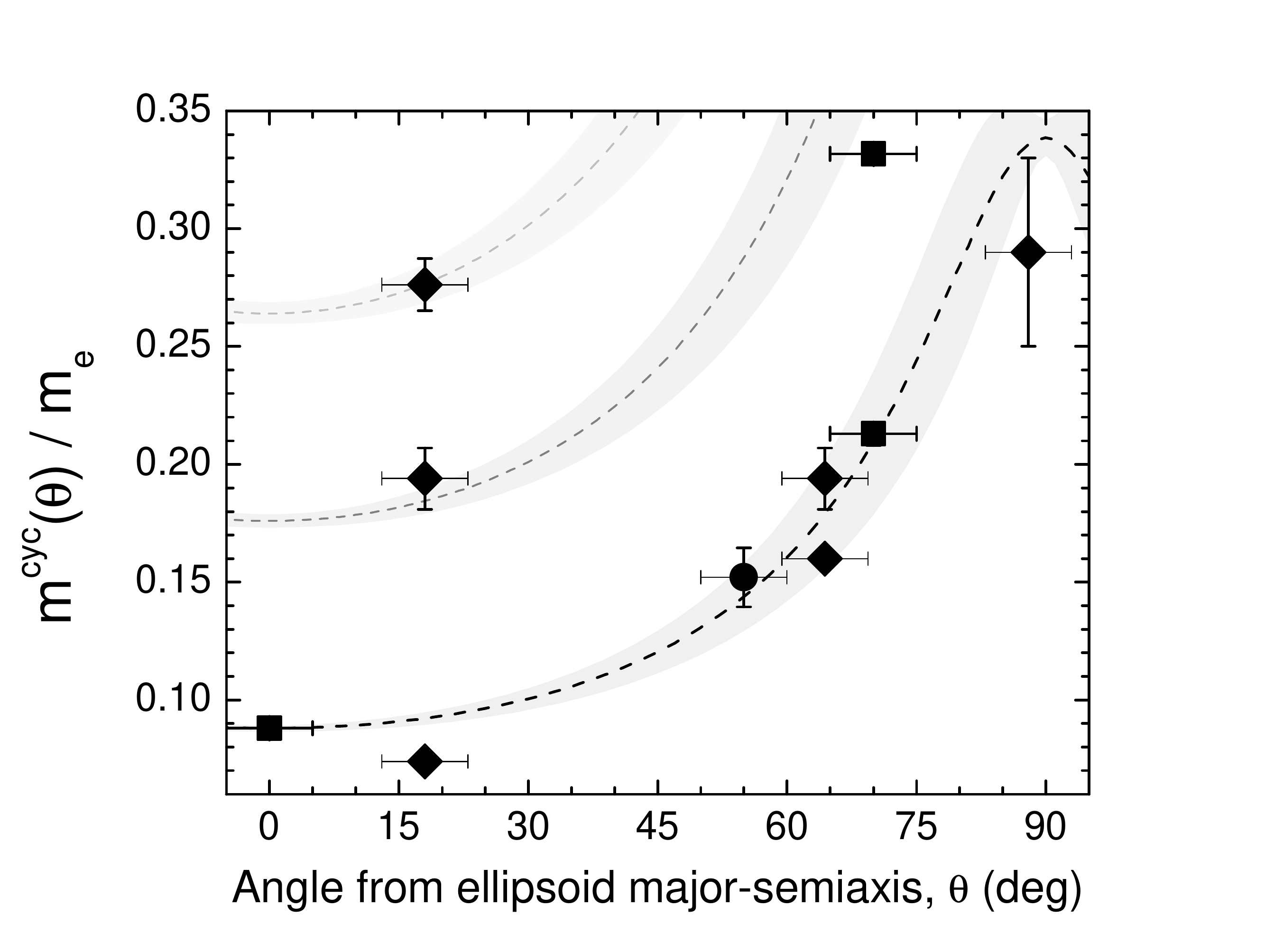} %\hspace{-0.7cm}
\caption{(Color online) Cyclotron effective mass, $m^{\text{cyc}}$, along different directions with respect to the (L-pocket) ellipsoid major semiaxis, for a Pb$_{1-x}$Na$_x$Te sample with $x=$0.24$\%$. The data points were obtained through fits to the LK-formula of the oscillating components of magnetoresistance along three different crystallographic directions: [111] (Fig. \ref{fig_Tdepall111}(b)), (100) (Fig. \ref{fig_Tdepall100}(d)), and 35$^{\circ}$ from [100] in the (110) plane (Fig. \ref{fig_Tdepall35}). The dashed lines represent the angle dependence of the cyclotron mass (fundamental and higher harmonics) for a perfect parabolic dispersion and perfect ellipsoidal model, as presented in Eqn. \ref{masscycangle}, and using an anisotropy parameter $K=14.3\pm0.4$ (which implies $m^{cyc}_{\|}/m^{cyc}_{\bot}=$3.78$\pm$0.05). The shadowed region around the dashed lines represents the error bar in $m^{cyc}(\theta)$ estimated from propagation of errors in $K$, $\theta$ and $m^{cyc}_{\bot}$.}\label{fig_mangle4505}
\end{figure}

\begin{figure}[!htbp]
\centering
%\hspace{-1cm}
\includegraphics[scale=0.3]{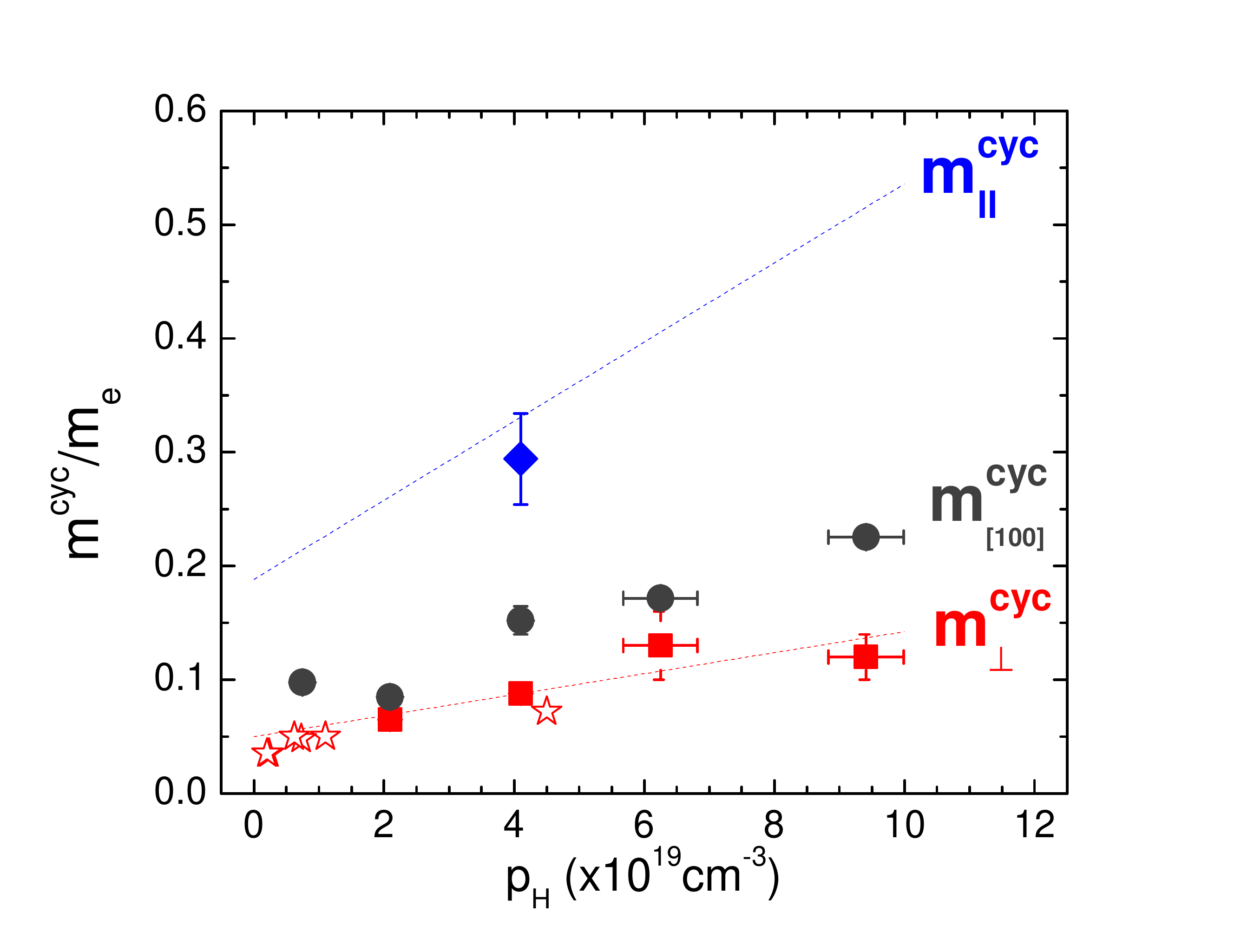} 
\caption{(Color online) Effective cyclotron mass, $m^{\text{cyc}}$, along three high symmetry directions for Pb$_{1-x}$Na$_x$Te samples, as a function of the Hall number. Cyclotron effective masses were determined through fitting  the curves in figures \ref{fig_Tdepall111}, \ref{fig_Tdepall35} and \ref{fig_Tdepall100} to the LK-formula in equation \ref{eq_LKmass}. $m_{\bot}^{cyc}$ is the cyclotron mass in the transverse direction of the L-pocket ellipsoid, or [111] direction; $m_{\|}^{cyc}$ is the cyclotron mass in the longitudinal direction of the L-pocket ellipsoid; and $m_{[100]}^{cyc}$ is the cyclotron mass in the [100] direction of the crystal lattice. The red dashed line represents a guide to the eye for the trend observed in the longitudinal cyclotron mass. The blue dashed-line is the trend expected for the longitudinal cyclotron mass given the anisotropy parameter of $m^{cyc}_{\|}/m^{cyc}_{\bot}=\sqrt{K}=3.78$.}\label{fig_massAll}
\end{figure}

As we presented in section \ref{sec_Tdep}, effective cyclotron masses along different high symmetry directions were obtained through direct fitting of the curves shown in Figs. \ref{fig_Tdepall111}, \ref{fig_Tdepall35} and \ref{fig_Tdepall100} to the LK-formula in equation \ref{eq_LKmass}. For all the Na compositions studied, the cyclotron masses along the transverse direction, $m_{\bot}^{cyc}$, and [100] direction, $m_{[100]}^{cyc}$, were determined through this method. Additionally, for samples with a Na concentration of $x=0.26\%$, the longitudinal cyclotron mass, $m_{||}^{cyc}$, was also found. Supplementary to these highly symmetric masses, others along less symmetric directions of the ellipsoid can be found from the different frequency terms in the measurements. Figure \ref{fig_mangle4505} shows the cyclotron effective masses found for all frequency terms taken into account in the LK fits of the $x=0.26\%$ sample (Figs. \ref{fig_Tdepall111}(b), Fig. \ref{fig_Tdepall35} and Fig. \ref{fig_Tdepall100}(c)), as a function of the angle from the L-pocket longitudinal direction. The corresponding angle for the mass of each frequency term, with respect to the longitudinal direction of the ellipsoids, was found by identifying each frequency in the angle dependence curves, such as that presented in Fig. \ref{fig_SdHall}(c). Fig. \ref{fig_mangle4505} also shows the expected angular dependence of the cyclotron effective mass (fundamental and higher harmonics) in a perfect ellipsoidal model (for more details, see appendix \ref{app_massan}), using the average $K$ value from Fig. \ref{fig_KNa} ($K=14.3\pm0.4$, which gives $m^{cyc}_{\|}/m^{cyc}_{\bot}=\sqrt{K}=3.78\pm0.05$). Most data points lie on this curve, confirming the good agreement of the topology of the Fermi surface with the perfect ellipsoidal model for this Na concentration.

In spite of the good agreement of the anisotropy of the cyclotron effective mass with the perfect ellipsoidal model, intriguingly, the masses are not constant throughout the band. Fig. \ref{fig_massAll} shows the evolution of the longitudinal, transverse and [100] direction cyclotron effective masses with carrier concentration. All of them show a monotonic increase with increasing carrier concentration, consistent with the predictions of the DFT band-structure calculations presented in section \ref{sec_DFTSdH}. Previous SdH measurements in p-type self-doped PbTe by Burke et al. \cite{burke2,burke1} ($p_H<1\times10^{19}$), and by Cuff et al. \cite{Cuff1} ($p_H<6\times10^{18}$), found a similar tendency for the transverse cyclotron mass. The observation of a varying effective mass with carrier concentration implies deviations from perfect parabolicity, starting from the top of the band.   

A Kane model dispersion relation has been proposed before to describe the valence band of PbTe \cite{Kanebook, Ravichbook, Salameh, Bilc1, Kong1}. In this model the non-parabolicity of the band is introduced as $E\longrightarrow\gamma(E)=E(1+E/E_g)$ in the dispersion relation, where $E_g$ is the band gap. For such a model, the longitudinal and transverse effective masses depend on energy in the same way \cite{Ravichbook}, implying that, although the effective masses evolve as the Fermi energy is changed, the band anisotropy parameter, $K=(A_{\|}/A_{\bot})^2=(m^{cyc}_{\|}/m^{cyc}_{\bot})^2$ is constant. Additionally, in this model, the constant energy surfaces for any Fermi energy are ellipsoids of revolution \cite{Ravichbook}, which is consistent with our observations for carrier concentrations up to $p=6.3\times10^{19}$cm$^{-3}$. The Kane model has been successful at describing the band structure near the gap of small band-gap semiconductors, for which the relevant Fermi energies are smaller than or of the same order as the band gap \cite{Kane1}. Our experimental results are in line with the predictions of the Kane model, ruling out other proposed models such as the Cohen model \cite{cohen1, Dixon1, Ravichbook}, at least for the low temperature regime.

\begin{figure}[!t]
\centering
\hspace{0.5cm}
\includegraphics[scale=0.3]{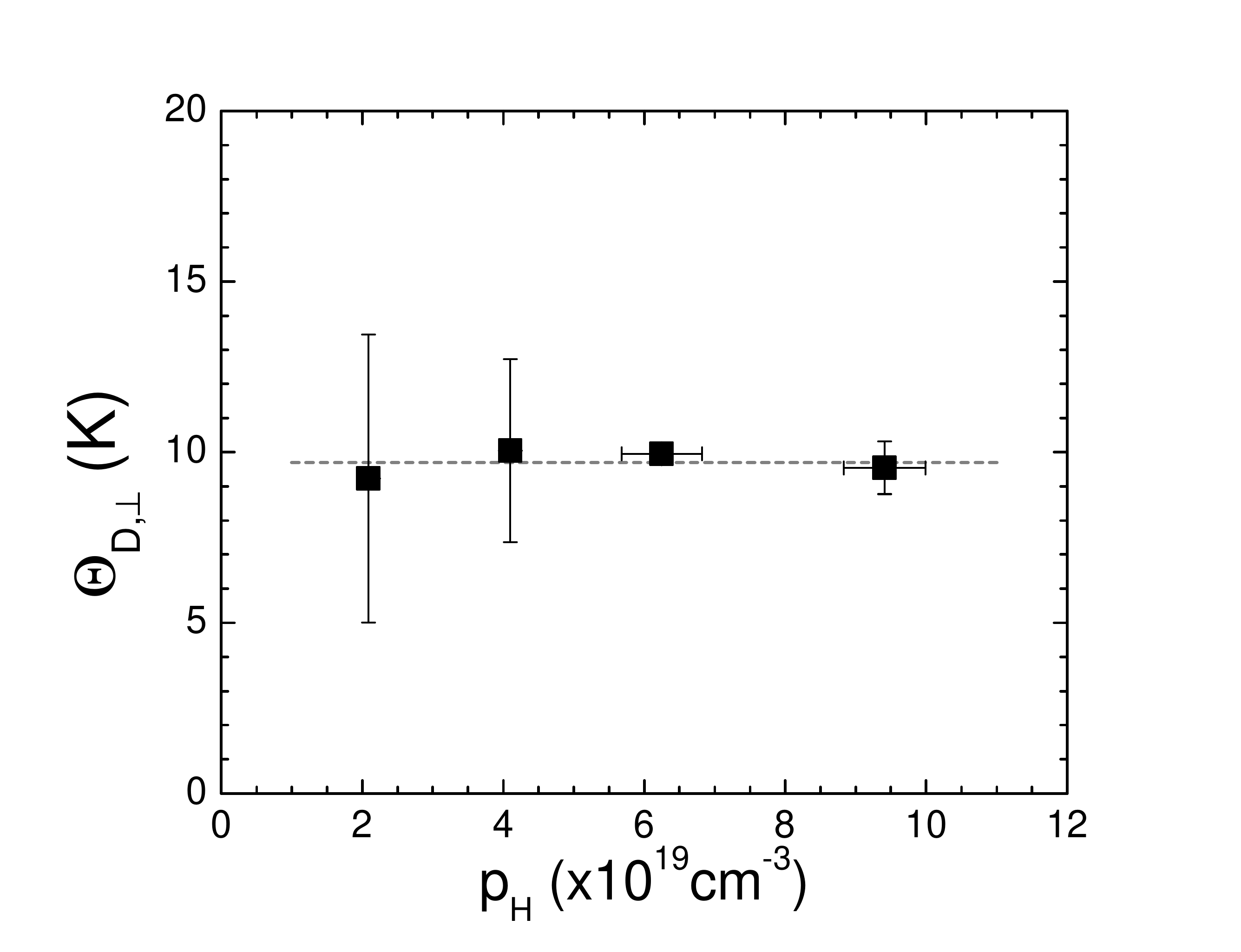} 
\caption{(Color online) Dingle temperature in the transverse direction, $\Theta_{D,\bot}$, obtained through fitting of the curves in Fig. \ref{fig_Tdepall111} to the LK formula in eq. \ref{eq_LKmass}, as a function of carrier concentration $p_H$. We find that the Dingle temperature is independent of carrier concentration, with a value of $\Theta_{D,\bot}$=(9.7$\pm$0.4) K, indicated by the dashed-gray line. This value of $\Theta_{D,\bot}$ results in a value of $\tau=$(0.125$\pm$0.005) ps for the carrier relaxation time along the transverse direction.}\label{fig_TDingle}
\end{figure}

Additional to the cyclotron effective masses, we have found the Dingle temperature in the transverse direction, $\Theta_{D,\bot}$, through a fitting of the data to the LK-formula, as presented in section \ref{sec_Tdep}. In contrast to the cyclotron mass, finding this quantity along directions other than the longitudinal axis of the L-pocket ellipses is challenging, given that the oscillatory part of magnetoresistance is dominated by the lowest frequency component. For this dominating part of the signal, the exponential damping in $1/H$ is the only one strong enough to result in a Dingle temperature as a strong fitting parameter. Fig. \ref{fig_TDingle} shows the Dingle temperature associated with the longitudinal direction, $\Theta_{D,\bot}$, as a function of carrier concentration. This quantity is constant for the range of concentrations studied, with an average value of $\Theta_{D,\bot}$=(9.7$\pm$0.4) K. This average value of $\Theta_{D,\bot}$ results in a value of the carrier relaxation time along the transverse direction of $\tau_{\bot}=\hbar/2\pi k_B\Theta_{D,\bot}=$(0.125$\pm$0.005) ps.

\section{Summary and Conclusions}

In summary, we have presented a numerical and experimental study of the low-temperature (1.3 K to 60 K) topology and properties of the Fermi surface of Pb$_{1-x}$Na$_x$Te, and its evolution with carrier concentration, for Na dopings up to $x$=0.62$\%$. We have found that:\\
(i) Although the band offset is underestimated by the DFT calculations, all the qualitative features of the evolution of the Fermi surface topology and effective mass are correctly predicted by our calculations. The underestimation of the band offset is related to the high sensitivity of the resulting band structure to variations of parameters in the calculation, such as lattice spacing or spin-orbit coupling. This fact is presumably related to the fact that PbTe is on the boundary between various competing structural (incipient ferroelectricity) and electronic (superconductivity and topological insulator) instabilities.\\
(ii) The Fermi surface of Pb$_{1-x}$Na$_x$Te up to a carrier concentration of $p=9.4\times 10^{19}$cm$^{-3}$ ($x=0.62\%$ - maximum studied) is formed solely by eight half ellipsoids at the L-points. The $\Sigma$-pockets predicted to contribute at such high carrier concentrations in our calculation and those of other groups \cite{Lin1,Dimmock1,Tung1,Martinez1,Parker}, are not observed. Additionally, the measured Hall number, and the Luttinger volume of the L-pockets calculated from our quantum oscillation measurements, match exactly, indicating that this is the only set of pockets that contribute to conduction in this compound at low temperatures.\\
(iii) The topology of the Fermi surface, formed by eight half pockets at the L-points, is well described by a perfect ellipsoidal model for carrier concentrations up to $p=6.3\times 10^{19}$cm$^{-3}$ ($x=0.4\%$). Deviations from perfect ellipsoidicity were resolved for the highest carrier concentration studied, $p=9.4\times 10^{19}$cm$^{-3}$ ($x=0.62\%$).\\
(iv) The anisotropy of the L-pockets is constant for the range of concentrations studied, and has an average value of $K=$14.3$\pm$0.4. \\
(v) The anisotropy of the cyclotron effective mass of the L-pockets follows the angular dependence expected in a perfect ellipsoidal model.\\
(vi) The effective cyclotron masses along all high symmetry directions increase monotonically with increasing carrier concentration, implying deviations from perfect parabolicity of the band. The observation of constant geometric and mass anisotropy with carrier concentration, but an increasing effective mass, is consistent with a Kane model of non-parabolic dispersion relation for the valence band of PbTe.

\begin{acknowledgments}
The high-field magnetoresistance measurements were performed at the National High Magnetic Field Laboratory (NHMFL), which is supported by NSF DMR-1157490 and the State of Florida. PGG, PW and IRF were supported by AFOSR Grant No. FA9550-09-1-0583. BS, MF and NAS acknowledge support from ETH Z\"urich, ERC Advanced Grant program (No. 291151), and the Swiss National Supercomputing Centre (CSCS) under project ID s307.

\end{acknowledgments}
	
%\bibliography{1_PbTe_paper_PRB_NadopingBiblio}\vspace{-0.8cm}

\begin{thebibliography}{57}%
\makeatletter
\providecommand \@ifxundefined [1]{%
 \@ifx{#1\undefined}
}%
\providecommand \@ifnum [1]{%
 \ifnum #1\expandafter \@firstoftwo
 \else \expandafter \@secondoftwo
 \fi
}%
\providecommand \@ifx [1]{%
 \ifx #1\expandafter \@firstoftwo
 \else \expandafter \@secondoftwo
 \fi
}%
\providecommand \natexlab [1]{#1}%
\providecommand \enquote  [1]{``#1''}%
\providecommand \bibnamefont  [1]{#1}%
\providecommand \bibfnamefont [1]{#1}%
\providecommand \citenamefont [1]{#1}%
\providecommand \href@noop [0]{\@secondoftwo}%
\providecommand \href [0]{\begingroup \@sanitize@url \@href}%
\providecommand \@href[1]{\@@startlink{#1}\@@href}%
\providecommand \@@href[1]{\endgroup#1\@@endlink}%
\providecommand \@sanitize@url [0]{\catcode `\\12\catcode `\$12\catcode
  `\&12\catcode `\#12\catcode `\^12\catcode `\_12\catcode `\%12\relax}%
\providecommand \@@startlink[1]{}%
\providecommand \@@endlink[0]{}%
\providecommand \url  [0]{\begingroup\@sanitize@url \@url }%
\providecommand \@url [1]{\endgroup\@href {#1}{\urlprefix }}%
\providecommand \urlprefix  [0]{URL }%
\providecommand \Eprint [0]{\href }%
\providecommand \doibase [0]{http://dx.doi.org/}%
\providecommand \selectlanguage [0]{\@gobble}%
\providecommand \bibinfo  [0]{\@secondoftwo}%
\providecommand \bibfield  [0]{\@secondoftwo}%
\providecommand \translation [1]{[#1]}%
\providecommand \BibitemOpen [0]{}%
\providecommand \bibitemStop [0]{}%
\providecommand \bibitemNoStop [0]{.\EOS\space}%
\providecommand \EOS [0]{\spacefactor3000\relax}%
\providecommand \BibitemShut  [1]{\csname bibitem#1\endcsname}%
\let\auto@bib@innerbib\@empty
%</preamble>
\bibitem [{\citenamefont {Ravich}\ \emph {et~al.}(1970)\citenamefont {Ravich},
  \citenamefont {Efimova},\ and\ \citenamefont {Smirnov}}]{Ravichbook2}%
  \BibitemOpen
  \bibfield  {author} {\bibinfo {author} {\bibfnamefont {Y.~I.}\ \bibnamefont
  {Ravich}}, \bibinfo {author} {\bibfnamefont {B.~A.}\ \bibnamefont {Efimova}},
  \ and\ \bibinfo {author} {\bibfnamefont {I.~A.}\ \bibnamefont {Smirnov}},\
  }\href@noop {} {\emph {\bibinfo {title} {Semiconducting Lead
  Chalcogenides}}}\ (\bibinfo  {publisher} {Plenum Press, New York},\ \bibinfo
  {year} {1970})\BibitemShut {NoStop}%
\bibitem [{\citenamefont {Dornhaus}\ \emph {et~al.}(1983)\citenamefont
  {Dornhaus}, \citenamefont {Nimtz},\ and\ \citenamefont
  {Schlicht}}]{Dornbook}%
  \BibitemOpen
  \bibfield  {author} {\bibinfo {author} {\bibfnamefont {R.}~\bibnamefont
  {Dornhaus}}, \bibinfo {author} {\bibfnamefont {G.}~\bibnamefont {Nimtz}}, \
  and\ \bibinfo {author} {\bibfnamefont {B.}~\bibnamefont {Schlicht}},\
  }\href@noop {} {\emph {\bibinfo {title} {Narrow-Gap Semiconductors, Springer
  Tracts Mod. Phys., vol. 98}}}\ (\bibinfo  {publisher} {Springer-Verlag, New
  York},\ \bibinfo {year} {1983})\BibitemShut {NoStop}%
\bibitem [{\citenamefont {Khokhlov}(2003)}]{Khokhlov}%
  \BibitemOpen
  \bibfield  {author} {\bibinfo {author} {\bibfnamefont {D.}~\bibnamefont
  {Khokhlov}},\ }\href@noop {} {\emph {\bibinfo {title} {Lead Chalcogenides:
  Physics and applications}}}\ (\bibinfo  {publisher} {Taylor and Francis
  books, New York},\ \bibinfo {year} {2003})\BibitemShut {NoStop}%
\bibitem [{\citenamefont {Pei}\ \emph {et~al.}(2011{\natexlab{a}})\citenamefont
  {Pei}, \citenamefont {{LaLonde}}, \citenamefont {Iwanaga},\ and\
  \citenamefont {Snyder}}]{Snyder2}%
  \BibitemOpen
  \bibfield  {author} {\bibinfo {author} {\bibfnamefont {Y.}~\bibnamefont
  {Pei}}, \bibinfo {author} {\bibfnamefont {A.}~\bibnamefont {{LaLonde}}},
  \bibinfo {author} {\bibfnamefont {S.}~\bibnamefont {Iwanaga}}, \ and\
  \bibinfo {author} {\bibfnamefont {G.~J.}\ \bibnamefont {Snyder}},\
  }\href@noop {} {\bibfield  {journal} {\bibinfo  {journal} {Energy Environ.
  Sci.}\ }\textbf {\bibinfo {volume} {4}},\ \bibinfo {pages} {2085} (\bibinfo
  {year} {2011}{\natexlab{a}})}\BibitemShut {NoStop}%
\bibitem [{\citenamefont {Pei}\ \emph {et~al.}(2011{\natexlab{b}})\citenamefont
  {Pei}, \citenamefont {Shi}, \citenamefont {LaLonde}, \citenamefont {Wang},
  \citenamefont {Chen},\ and\ \citenamefont {Snyder}}]{Snyder1}%
  \BibitemOpen
  \bibfield  {author} {\bibinfo {author} {\bibfnamefont {Y.}~\bibnamefont
  {Pei}}, \bibinfo {author} {\bibfnamefont {X.}~\bibnamefont {Shi}}, \bibinfo
  {author} {\bibfnamefont {A.}~\bibnamefont {LaLonde}}, \bibinfo {author}
  {\bibfnamefont {H.}~\bibnamefont {Wang}}, \bibinfo {author} {\bibfnamefont
  {L.}~\bibnamefont {Chen}}, \ and\ \bibinfo {author} {\bibfnamefont {G.~J.}\
  \bibnamefont {Snyder}},\ }\href@noop {} {\bibfield  {journal} {\bibinfo
  {journal} {Nature}\ }\textbf {\bibinfo {volume} {473}},\ \bibinfo {pages}
  {66} (\bibinfo {year} {2011}{\natexlab{b}})}\BibitemShut {NoStop}%
\bibitem [{\citenamefont {Kanatzidis}(2010)}]{Kanatzidis1}%
  \BibitemOpen
  \bibfield  {author} {\bibinfo {author} {\bibfnamefont {M.~G.}\ \bibnamefont
  {Kanatzidis}},\ }\href@noop {} {\bibfield  {journal} {\bibinfo  {journal}
  {Energy Environ. Sci.}\ }\textbf {\bibinfo {volume} {22}},\ \bibinfo {pages}
  {648} (\bibinfo {year} {2010})}\BibitemShut {NoStop}%
\bibitem [{\citenamefont {Heremans}\ \emph {et~al.}(2008)\citenamefont
  {Heremans}, \citenamefont {Jovovic}, \citenamefont {Toberer}, \citenamefont
  {Saramat}, \citenamefont {Kurosaki}, \citenamefont {Charoenphakdee},
  \citenamefont {Yamanaka},\ and\ \citenamefont {Snyder}}]{Snyder3}%
  \BibitemOpen
  \bibfield  {author} {\bibinfo {author} {\bibfnamefont {J.~P.}\ \bibnamefont
  {Heremans}}, \bibinfo {author} {\bibfnamefont {V.}~\bibnamefont {Jovovic}},
  \bibinfo {author} {\bibfnamefont {E.~S.}\ \bibnamefont {Toberer}}, \bibinfo
  {author} {\bibfnamefont {A.}~\bibnamefont {Saramat}}, \bibinfo {author}
  {\bibfnamefont {K.}~\bibnamefont {Kurosaki}}, \bibinfo {author}
  {\bibfnamefont {A.}~\bibnamefont {Charoenphakdee}}, \bibinfo {author}
  {\bibfnamefont {S.}~\bibnamefont {Yamanaka}}, \ and\ \bibinfo {author}
  {\bibfnamefont {G.~J.}\ \bibnamefont {Snyder}},\ }\href@noop {} {\bibfield
  {journal} {\bibinfo  {journal} {Science}\ }\textbf {\bibinfo {volume}
  {321}},\ \bibinfo {pages} {554} (\bibinfo {year} {2008})}\BibitemShut
  {NoStop}%
\bibitem [{\citenamefont {Kaidanov}\ and\ \citenamefont
  {Ravich}(1985)}]{Ravich}%
  \BibitemOpen
  \bibfield  {author} {\bibinfo {author} {\bibfnamefont {V.~I.}\ \bibnamefont
  {Kaidanov}}\ and\ \bibinfo {author} {\bibfnamefont {Y.~I.}\ \bibnamefont
  {Ravich}},\ }\href@noop {} {\bibfield  {journal} {\bibinfo  {journal} {Sov.
  Phys. Usp.}\ }\textbf {\bibinfo {volume} {28}},\ \bibinfo {pages} {31}
  (\bibinfo {year} {1985})}\BibitemShut {NoStop}%
\bibitem [{\citenamefont {Lewis}(1970)}]{Lewis1}%
  \BibitemOpen
  \bibfield  {author} {\bibinfo {author} {\bibfnamefont {J.~E.}\ \bibnamefont
  {Lewis}},\ }\href@noop {} {\bibfield  {journal} {\bibinfo  {journal} {Phys.
  Stat. sol (b)}\ }\textbf {\bibinfo {volume} {42}},\ \bibinfo {pages} {K97}
  (\bibinfo {year} {1970})}\BibitemShut {NoStop}%
\bibitem [{\citenamefont {Matsushita}\ \emph {et~al.}(2005)\citenamefont
  {Matsushita}, \citenamefont {Bluhm}, \citenamefont {Geballe},\ and\
  \citenamefont {Fisher}}]{Yana1}%
  \BibitemOpen
  \bibfield  {author} {\bibinfo {author} {\bibfnamefont {Y.}~\bibnamefont
  {Matsushita}}, \bibinfo {author} {\bibfnamefont {H.}~\bibnamefont {Bluhm}},
  \bibinfo {author} {\bibfnamefont {T.~H.}\ \bibnamefont {Geballe}}, \ and\
  \bibinfo {author} {\bibfnamefont {I.~R.}\ \bibnamefont {Fisher}},\
  }\href@noop {} {\bibfield  {journal} {\bibinfo  {journal} {Phys. \ Rev.
  Lett.}\ }\textbf {\bibinfo {volume} {94}},\ \bibinfo {pages} {157002}
  (\bibinfo {year} {2005})}\BibitemShut {NoStop}%
\bibitem [{\citenamefont {Matsushita}\ \emph {et~al.}(2006)\citenamefont
  {Matsushita}, \citenamefont {Sommer}, \citenamefont {Geballe},\ and\
  \citenamefont {Fisher}}]{Yana2}%
  \BibitemOpen
  \bibfield  {author} {\bibinfo {author} {\bibfnamefont {Y.}~\bibnamefont
  {Matsushita}}, \bibinfo {author} {\bibfnamefont {P.~A. W. A.~T.}\
  \bibnamefont {Sommer}}, \bibinfo {author} {\bibfnamefont {T.~H.}\
  \bibnamefont {Geballe}}, \ and\ \bibinfo {author} {\bibfnamefont {I.~R.}\
  \bibnamefont {Fisher}},\ }\href@noop {} {\bibfield  {journal} {\bibinfo
  {journal} {Phys. \ Rev. B}\ }\textbf {\bibinfo {volume} {74}},\ \bibinfo
  {pages} {134512} (\bibinfo {year} {2006})}\BibitemShut {NoStop}%
\bibitem [{\citenamefont {Dzero}\ and\ \citenamefont
  {Schmalian}(2005)}]{DZero}%
  \BibitemOpen
  \bibfield  {author} {\bibinfo {author} {\bibfnamefont {M.}~\bibnamefont
  {Dzero}}\ and\ \bibinfo {author} {\bibfnamefont {J.}~\bibnamefont
  {Schmalian}},\ }\href@noop {} {\bibfield  {journal} {\bibinfo  {journal}
  {Phys. \ Rev. Lett.}\ }\textbf {\bibinfo {volume} {94}},\ \bibinfo {pages}
  {157003} (\bibinfo {year} {2005})}\BibitemShut {NoStop}%
\bibitem [{\citenamefont {Parker}\ \emph {et~al.}(2013)\citenamefont {Parker},
  \citenamefont {Chen},\ and\ \citenamefont {Singh}}]{Parker}%
  \BibitemOpen
  \bibfield  {author} {\bibinfo {author} {\bibfnamefont {D.}~\bibnamefont
  {Parker}}, \bibinfo {author} {\bibfnamefont {X.}~\bibnamefont {Chen}}, \ and\
  \bibinfo {author} {\bibfnamefont {D.~J.}\ \bibnamefont {Singh}},\ }\href@noop
  {} {\bibfield  {journal} {\bibinfo  {journal} {Phys. \ Rev. Lett.}\ }\textbf
  {\bibinfo {volume} {110}},\ \bibinfo {pages} {146601} (\bibinfo {year}
  {2013})}\BibitemShut {NoStop}%
\bibitem [{\citenamefont {Burke}\ \emph {et~al.}(1970)\citenamefont {Burke},
  \citenamefont {Houston},\ and\ \citenamefont {Savage}}]{burke2}%
  \BibitemOpen
  \bibfield  {author} {\bibinfo {author} {\bibfnamefont {J.~R.}\ \bibnamefont
  {Burke}}, \bibinfo {author} {\bibfnamefont {B.}~\bibnamefont {Houston}}, \
  and\ \bibinfo {author} {\bibfnamefont {H.~T.}\ \bibnamefont {Savage}},\
  }\href@noop {} {\bibfield  {journal} {\bibinfo  {journal} {Phys.\ Rev. B}\
  }\textbf {\bibinfo {volume} {2}},\ \bibinfo {pages} {1977} (\bibinfo {year}
  {1970})}\BibitemShut {NoStop}%
\bibitem [{\citenamefont {Jensen}\ \emph {et~al.}(1978)\citenamefont {Jensen},
  \citenamefont {Houston},\ and\ \citenamefont {Burke}}]{burke1}%
  \BibitemOpen
  \bibfield  {author} {\bibinfo {author} {\bibfnamefont {J.~D.}\ \bibnamefont
  {Jensen}}, \bibinfo {author} {\bibfnamefont {B.}~\bibnamefont {Houston}}, \
  and\ \bibinfo {author} {\bibfnamefont {J.~R.}\ \bibnamefont {Burke}},\
  }\href@noop {} {\bibfield  {journal} {\bibinfo  {journal} {Phys.\ Rev. B}\
  }\textbf {\bibinfo {volume} {18}},\ \bibinfo {pages} {5567} (\bibinfo {year}
  {1978})}\BibitemShut {NoStop}%
\bibitem [{\citenamefont {Sitter}\ \emph {et~al.}(1977)\citenamefont {Sitter},
  \citenamefont {Lischka},\ and\ \citenamefont {Heinrich}}]{Sitter1}%
  \BibitemOpen
  \bibfield  {author} {\bibinfo {author} {\bibfnamefont {H.}~\bibnamefont
  {Sitter}}, \bibinfo {author} {\bibfnamefont {K.}~\bibnamefont {Lischka}}, \
  and\ \bibinfo {author} {\bibfnamefont {H.}~\bibnamefont {Heinrich}},\
  }\href@noop {} {\bibfield  {journal} {\bibinfo  {journal} {Phys. Rev. B}\
  }\textbf {\bibinfo {volume} {16}},\ \bibinfo {pages} {680} (\bibinfo {year}
  {1977})}\BibitemShut {NoStop}%
\bibitem [{\citenamefont {An}\ \emph {et~al.}(2008)\citenamefont {An},
  \citenamefont {Subedi},\ and\ \citenamefont {Singh}}]{an2008}%
  \BibitemOpen
  \bibfield  {author} {\bibinfo {author} {\bibfnamefont {J.}~\bibnamefont
  {An}}, \bibinfo {author} {\bibfnamefont {A.}~\bibnamefont {Subedi}}, \ and\
  \bibinfo {author} {\bibfnamefont {D.}~\bibnamefont {Singh}},\ }\href
  {\doibase 10.1016/j.ssc.2008.09.027} {\bibfield  {journal} {\bibinfo
  {journal} {Solid State Commun.}\ }\textbf {\bibinfo {volume} {148}},\
  \bibinfo {pages} {417} (\bibinfo {year} {2008})}\BibitemShut {NoStop}%
\bibitem [{\citenamefont {Zhang}\ \emph {et~al.}(2009)\citenamefont {Zhang},
  \citenamefont {Ke}, \citenamefont {Chen}, \citenamefont {Yang},\ and\
  \citenamefont {Kent}}]{zhang2009}%
  \BibitemOpen
  \bibfield  {author} {\bibinfo {author} {\bibfnamefont {Y.}~\bibnamefont
  {Zhang}}, \bibinfo {author} {\bibfnamefont {X.}~\bibnamefont {Ke}}, \bibinfo
  {author} {\bibfnamefont {C.}~\bibnamefont {Chen}}, \bibinfo {author}
  {\bibfnamefont {J.}~\bibnamefont {Yang}}, \ and\ \bibinfo {author}
  {\bibfnamefont {P.}~\bibnamefont {Kent}},\ }\href {\doibase
  10.1103/PhysRevB.80.024304} {\bibfield  {journal} {\bibinfo  {journal} {Phys.
  Rev. B}\ }\textbf {\bibinfo {volume} {80}},\ \bibinfo {pages} {024304}
  (\bibinfo {year} {2009})}\BibitemShut {NoStop}%
\bibitem [{\citenamefont {Bo{\v{z}}in}\ \emph {et~al.}(2010)\citenamefont
  {Bo{\v{z}}in}, \citenamefont {Malliakas}, \citenamefont {Souvatzis},
  \citenamefont {Proffen}, \citenamefont {Spaldin}, \citenamefont
  {Kanatzidis},\ and\ \citenamefont {Billinge}}]{bozin2010}%
  \BibitemOpen
  \bibfield  {author} {\bibinfo {author} {\bibfnamefont {E.~S.}\ \bibnamefont
  {Bo{\v{z}}in}}, \bibinfo {author} {\bibfnamefont {C.~D.}\ \bibnamefont
  {Malliakas}}, \bibinfo {author} {\bibfnamefont {P.}~\bibnamefont
  {Souvatzis}}, \bibinfo {author} {\bibfnamefont {T.}~\bibnamefont {Proffen}},
  \bibinfo {author} {\bibfnamefont {N.~a.}\ \bibnamefont {Spaldin}}, \bibinfo
  {author} {\bibfnamefont {M.~G.}\ \bibnamefont {Kanatzidis}}, \ and\ \bibinfo
  {author} {\bibfnamefont {S.~J.~L.}\ \bibnamefont {Billinge}},\ }\href
  {\doibase 10.1126/science.1192759} {\bibfield  {journal} {\bibinfo  {journal}
  {Science}\ }\textbf {\bibinfo {volume} {330}},\ \bibinfo {pages} {1660}
  (\bibinfo {year} {2010})}\BibitemShut {NoStop}%
\bibitem [{\citenamefont {Erickson}\ \emph {et~al.}(2009)\citenamefont
  {Erickson}, \citenamefont {Chu}, \citenamefont {Toney}, \citenamefont
  {Geballe},\ and\ \citenamefont {Fisher}}]{Ann2}%
  \BibitemOpen
  \bibfield  {author} {\bibinfo {author} {\bibfnamefont {A.~S.}\ \bibnamefont
  {Erickson}}, \bibinfo {author} {\bibfnamefont {J.~H.}\ \bibnamefont {Chu}},
  \bibinfo {author} {\bibfnamefont {M.~F.}\ \bibnamefont {Toney}}, \bibinfo
  {author} {\bibfnamefont {T.~H.}\ \bibnamefont {Geballe}}, \ and\ \bibinfo
  {author} {\bibfnamefont {I.~R.}\ \bibnamefont {Fisher}},\ }\href@noop {}
  {\bibfield  {journal} {\bibinfo  {journal} {Phys. Rev. B}\ }\textbf {\bibinfo
  {volume} {79}},\ \bibinfo {pages} {024520} (\bibinfo {year}
  {2009})}\BibitemShut {NoStop}%
\bibitem [{\citenamefont {Barone}\ \emph
  {et~al.}(2013{\natexlab{a}})\citenamefont {Barone}, \citenamefont {Rauch},
  \citenamefont {{Di Sante}}, \citenamefont {Henk}, \citenamefont {Mertig},\
  and\ \citenamefont {Picozzi}}]{barone2013}%
  \BibitemOpen
  \bibfield  {author} {\bibinfo {author} {\bibfnamefont {P.}~\bibnamefont
  {Barone}}, \bibinfo {author} {\bibfnamefont {T.}~\bibnamefont {Rauch}},
  \bibinfo {author} {\bibfnamefont {D.}~\bibnamefont {{Di Sante}}}, \bibinfo
  {author} {\bibfnamefont {J.}~\bibnamefont {Henk}}, \bibinfo {author}
  {\bibfnamefont {I.}~\bibnamefont {Mertig}}, \ and\ \bibinfo {author}
  {\bibfnamefont {S.}~\bibnamefont {Picozzi}},\ }\href {\doibase
  10.1103/PhysRevB.88.045207} {\bibfield  {journal} {\bibinfo  {journal} {Phys.
  Rev. B}\ }\textbf {\bibinfo {volume} {88}},\ \bibinfo {pages} {045207}
  (\bibinfo {year} {2013}{\natexlab{a}})}\BibitemShut {NoStop}%
\bibitem [{\citenamefont {Barone}\ \emph
  {et~al.}(2013{\natexlab{b}})\citenamefont {Barone}, \citenamefont {{Di
  Sante}},\ and\ \citenamefont {Picozzi}}]{barone2013a}%
  \BibitemOpen
  \bibfield  {author} {\bibinfo {author} {\bibfnamefont {P.}~\bibnamefont
  {Barone}}, \bibinfo {author} {\bibfnamefont {D.}~\bibnamefont {{Di Sante}}},
  \ and\ \bibinfo {author} {\bibfnamefont {S.}~\bibnamefont {Picozzi}},\ }\href
  {\doibase 10.1002/pssr.201308154} {\bibfield  {journal} {\bibinfo  {journal}
  {Phys. status solidi - Rapid Res. Lett.}\ }\textbf {\bibinfo {volume} {7}},\
  \bibinfo {pages} {1102} (\bibinfo {year} {2013}{\natexlab{b}})}\BibitemShut
  {NoStop}%
\bibitem [{\citenamefont {Bl\"ochl}(1994)}]{bloechl1994}%
  \BibitemOpen
  \bibfield  {author} {\bibinfo {author} {\bibfnamefont {P.~E.}\ \bibnamefont
  {Bl\"ochl}},\ }\href {\doibase 10.1103/PhysRevB.50.17953} {\bibfield
  {journal} {\bibinfo  {journal} {Phys. Rev. B}\ }\textbf {\bibinfo {volume}
  {50}},\ \bibinfo {pages} {17953} (\bibinfo {year} {1994})}\BibitemShut
  {NoStop}%
\bibitem [{\citenamefont {Kresse}\ and\ \citenamefont
  {Joubert}(1999)}]{kresse1999}%
  \BibitemOpen
  \bibfield  {author} {\bibinfo {author} {\bibfnamefont {G.}~\bibnamefont
  {Kresse}}\ and\ \bibinfo {author} {\bibfnamefont {D.}~\bibnamefont
  {Joubert}},\ }\href {\doibase 10.1103/PhysRevB.59.1758} {\bibfield  {journal}
  {\bibinfo  {journal} {Phys. Rev. B}\ }\textbf {\bibinfo {volume} {59}},\
  \bibinfo {pages} {1758} (\bibinfo {year} {1999})}\BibitemShut {NoStop}%
\bibitem [{\citenamefont {Kresse}\ and\ \citenamefont
  {Furthm\"{u}ller}(1996)}]{kresse1996}%
  \BibitemOpen
  \bibfield  {author} {\bibinfo {author} {\bibfnamefont {G.}~\bibnamefont
  {Kresse}}\ and\ \bibinfo {author} {\bibfnamefont {J.}~\bibnamefont
  {Furthm\"{u}ller}},\ }\href {\doibase 10.1103/PhysRevB.54.11169} {\bibfield
  {journal} {\bibinfo  {journal} {Phys. Rev. B}\ }\textbf {\bibinfo {volume}
  {54}},\ \bibinfo {pages} {11169} (\bibinfo {year} {1996})}\BibitemShut
  {NoStop}%
\bibitem [{\citenamefont {Perdew}\ and\ \citenamefont
  {Zunger}(1981)}]{perdew1981}%
  \BibitemOpen
  \bibfield  {author} {\bibinfo {author} {\bibfnamefont {J.~P.}\ \bibnamefont
  {Perdew}}\ and\ \bibinfo {author} {\bibfnamefont {A.}~\bibnamefont
  {Zunger}},\ }\href {\doibase 10.1103/PhysRevB.23.5048} {\bibfield  {journal}
  {\bibinfo  {journal} {Phys. Rev. B}\ }\textbf {\bibinfo {volume} {23}},\
  \bibinfo {pages} {5048} (\bibinfo {year} {1981})}\BibitemShut {NoStop}%
\bibitem [{\citenamefont {Perdew}\ \emph {et~al.}(1996)\citenamefont {Perdew},
  \citenamefont {Burke},\ and\ \citenamefont {Ernzerhof}}]{perdew1996}%
  \BibitemOpen
  \bibfield  {author} {\bibinfo {author} {\bibfnamefont {J.~P.}\ \bibnamefont
  {Perdew}}, \bibinfo {author} {\bibfnamefont {K.}~\bibnamefont {Burke}}, \
  and\ \bibinfo {author} {\bibfnamefont {M.}~\bibnamefont {Ernzerhof}},\ }\href
  {\doibase 10.1103/PhysRevLett.77.3865} {\bibfield  {journal} {\bibinfo
  {journal} {Phys. Rev. Lett.}\ }\textbf {\bibinfo {volume} {77}},\ \bibinfo
  {pages} {3865} (\bibinfo {year} {1996})}\BibitemShut {NoStop}%
\bibitem [{\citenamefont {Perdew}\ \emph {et~al.}(2008)\citenamefont {Perdew},
  \citenamefont {Ruzsinszky}, \citenamefont {Csonka}, \citenamefont {Vydrov},
  \citenamefont {Scuseria}, \citenamefont {Constantin}, \citenamefont {Zhou},\
  and\ \citenamefont {Burke}}]{perdew2008}%
  \BibitemOpen
  \bibfield  {author} {\bibinfo {author} {\bibfnamefont {J.~P.}\ \bibnamefont
  {Perdew}}, \bibinfo {author} {\bibfnamefont {A.}~\bibnamefont {Ruzsinszky}},
  \bibinfo {author} {\bibfnamefont {G.~I.}\ \bibnamefont {Csonka}}, \bibinfo
  {author} {\bibfnamefont {O.~A.}\ \bibnamefont {Vydrov}}, \bibinfo {author}
  {\bibfnamefont {G.~E.}\ \bibnamefont {Scuseria}}, \bibinfo {author}
  {\bibfnamefont {L.~A.}\ \bibnamefont {Constantin}}, \bibinfo {author}
  {\bibfnamefont {X.}~\bibnamefont {Zhou}}, \ and\ \bibinfo {author}
  {\bibfnamefont {K.}~\bibnamefont {Burke}},\ }\href {\doibase
  10.1103/PhysRevLett.100.136406} {\bibfield  {journal} {\bibinfo  {journal}
  {Phys. Rev. Lett.}\ }\textbf {\bibinfo {volume} {100}},\ \bibinfo {pages}
  {136406} (\bibinfo {year} {2008})}\BibitemShut {NoStop}%
\bibitem [{\citenamefont {Hummer}\ \emph {et~al.}(2007)\citenamefont {Hummer},
  \citenamefont {Gr{\"{u}}neis},\ and\ \citenamefont {Kresse}}]{hummer2007}%
  \BibitemOpen
  \bibfield  {author} {\bibinfo {author} {\bibfnamefont {K.}~\bibnamefont
  {Hummer}}, \bibinfo {author} {\bibfnamefont {A.}~\bibnamefont
  {Gr{\"{u}}neis}}, \ and\ \bibinfo {author} {\bibfnamefont {G.}~\bibnamefont
  {Kresse}},\ }\href {\doibase 10.1103/PhysRevB.75.195211} {\bibfield
  {journal} {\bibinfo  {journal} {Phys. Rev. B}\ }\textbf {\bibinfo {volume}
  {75}},\ \bibinfo {pages} {195211} (\bibinfo {year} {2007})}\BibitemShut
  {NoStop}%
\bibitem [{\citenamefont {Rourke}\ and\ \citenamefont
  {Julian}(2012)}]{rourke2012}%
  \BibitemOpen
  \bibfield  {author} {\bibinfo {author} {\bibfnamefont {P.~M.~C.}\
  \bibnamefont {Rourke}}\ and\ \bibinfo {author} {\bibfnamefont {S.~R.}\
  \bibnamefont {Julian}},\ }\href {\doibase 10.1016/j.cpc.2011.10.015}
  {\bibfield  {journal} {\bibinfo  {journal} {Comput. Phys. Commun.}\ }\textbf
  {\bibinfo {volume} {183}},\ \bibinfo {pages} {324} (\bibinfo {year}
  {2012})}\BibitemShut {NoStop}%
\bibitem [{\citenamefont {Ahmad}\ \emph
  {et~al.}(2006{\natexlab{a}})\citenamefont {Ahmad}, \citenamefont {Hoang},\
  and\ \citenamefont {Mahanti}}]{ahmad2006}%
  \BibitemOpen
  \bibfield  {author} {\bibinfo {author} {\bibfnamefont {S.}~\bibnamefont
  {Ahmad}}, \bibinfo {author} {\bibfnamefont {K.}~\bibnamefont {Hoang}}, \ and\
  \bibinfo {author} {\bibfnamefont {S.~D.}\ \bibnamefont {Mahanti}},\ }\href
  {\doibase 10.1103/PhysRevLett.96.056403} {\bibfield  {journal} {\bibinfo
  {journal} {Phys. Rev. Lett.}\ }\textbf {\bibinfo {volume} {96}},\ \bibinfo
  {pages} {056403} (\bibinfo {year} {2006}{\natexlab{a}})}\BibitemShut
  {NoStop}%
\bibitem [{\citenamefont {Hase}\ and\ \citenamefont
  {Yanagisawa}(2006)}]{hase2006}%
  \BibitemOpen
  \bibfield  {author} {\bibinfo {author} {\bibfnamefont {I.}~\bibnamefont
  {Hase}}\ and\ \bibinfo {author} {\bibfnamefont {T.}~\bibnamefont
  {Yanagisawa}},\ }\href {\doibase 10.1016/j.physc.2006.03.078} {\bibfield
  {journal} {\bibinfo  {journal} {Phys. C Supercond.}\ }\textbf {\bibinfo
  {volume} {445-448}},\ \bibinfo {pages} {61} (\bibinfo {year}
  {2006})}\BibitemShut {NoStop}%
\bibitem [{\citenamefont {Ahmad}\ \emph
  {et~al.}(2006{\natexlab{b}})\citenamefont {Ahmad}, \citenamefont {Mahanti},
  \citenamefont {Hoang},\ and\ \citenamefont {Kanatzidis}}]{Salameh}%
  \BibitemOpen
  \bibfield  {author} {\bibinfo {author} {\bibfnamefont {S.}~\bibnamefont
  {Ahmad}}, \bibinfo {author} {\bibfnamefont {S.~D.}\ \bibnamefont {Mahanti}},
  \bibinfo {author} {\bibfnamefont {K.}~\bibnamefont {Hoang}}, \ and\ \bibinfo
  {author} {\bibfnamefont {M.~G.}\ \bibnamefont {Kanatzidis}},\ }\href@noop {}
  {\bibfield  {journal} {\bibinfo  {journal} {Phys.\ Rev. B}\ }\textbf
  {\bibinfo {volume} {74}},\ \bibinfo {pages} {155205} (\bibinfo {year}
  {2006}{\natexlab{b}})}\BibitemShut {NoStop}%
\bibitem [{\citenamefont {Xiong}\ \emph {et~al.}(2010)\citenamefont {Xiong},
  \citenamefont {Lee}, \citenamefont {Gupta}, \citenamefont {Wang},
  \citenamefont {Gnade},\ and\ \citenamefont {Cho}}]{Cho1}%
  \BibitemOpen
  \bibfield  {author} {\bibinfo {author} {\bibfnamefont {K.}~\bibnamefont
  {Xiong}}, \bibinfo {author} {\bibfnamefont {G.}~\bibnamefont {Lee}}, \bibinfo
  {author} {\bibfnamefont {R.~P.}\ \bibnamefont {Gupta}}, \bibinfo {author}
  {\bibfnamefont {W.}~\bibnamefont {Wang}}, \bibinfo {author} {\bibfnamefont
  {B.~E.}\ \bibnamefont {Gnade}}, \ and\ \bibinfo {author} {\bibfnamefont
  {K.}~\bibnamefont {Cho}},\ }\href@noop {} {\bibfield  {journal} {\bibinfo
  {journal} {J. Phys. D: Appl. Phys.}\ }\textbf {\bibinfo {volume} {43}},\
  \bibinfo {pages} {405403} (\bibinfo {year} {2010})}\BibitemShut {NoStop}%
\bibitem [{\citenamefont {Takagiwa}\ \emph {et~al.}(2013)\citenamefont
  {Takagiwa}, \citenamefont {Pei}, \citenamefont {Pomrehn},\ and\ \citenamefont
  {{Jeffrey Snyder}}}]{takagiwa2013}%
  \BibitemOpen
  \bibfield  {author} {\bibinfo {author} {\bibfnamefont {Y.}~\bibnamefont
  {Takagiwa}}, \bibinfo {author} {\bibfnamefont {Y.}~\bibnamefont {Pei}},
  \bibinfo {author} {\bibfnamefont {G.}~\bibnamefont {Pomrehn}}, \ and\
  \bibinfo {author} {\bibfnamefont {G.}~\bibnamefont {{Jeffrey Snyder}}},\
  }\href {\doibase 10.1063/1.4809545} {\bibfield  {journal} {\bibinfo
  {journal} {APL Mater.}\ }\textbf {\bibinfo {volume} {1}},\ \bibinfo {pages}
  {011101} (\bibinfo {year} {2013})}\BibitemShut {NoStop}%
\bibitem [{\citenamefont {Lee}\ and\ \citenamefont {Mahanti}(2012)}]{lee2012}%
  \BibitemOpen
  \bibfield  {author} {\bibinfo {author} {\bibfnamefont {M.-S.}\ \bibnamefont
  {Lee}}\ and\ \bibinfo {author} {\bibfnamefont {S.~D.}\ \bibnamefont
  {Mahanti}},\ }\href {\doibase 10.1103/PhysRevB.85.165149} {\bibfield
  {journal} {\bibinfo  {journal} {Phys. Rev. B}\ }\textbf {\bibinfo {volume}
  {85}},\ \bibinfo {pages} {165149} (\bibinfo {year} {2012})}\BibitemShut
  {NoStop}%
\bibitem [{\citenamefont {Hoang}\ \emph {et~al.}(2010)\citenamefont {Hoang},
  \citenamefont {Mahanti},\ and\ \citenamefont {Kanatzidis}}]{hoang2010}%
  \BibitemOpen
  \bibfield  {author} {\bibinfo {author} {\bibfnamefont {K.}~\bibnamefont
  {Hoang}}, \bibinfo {author} {\bibfnamefont {S.~D.}\ \bibnamefont {Mahanti}},
  \ and\ \bibinfo {author} {\bibfnamefont {M.~G.}\ \bibnamefont {Kanatzidis}},\
  }\href {\doibase 10.1103/PhysRevB.81.115106} {\bibfield  {journal} {\bibinfo
  {journal} {Phys. Rev. B}\ }\textbf {\bibinfo {volume} {81}},\ \bibinfo
  {pages} {115106} (\bibinfo {year} {2010})}\BibitemShut {NoStop}%
\bibitem [{\citenamefont {Venkatapathi}\ \emph {et~al.}(2014)\citenamefont
  {Venkatapathi}, \citenamefont {Dong},\ and\ \citenamefont
  {Hin}}]{venkatapathi2014}%
  \BibitemOpen
  \bibfield  {author} {\bibinfo {author} {\bibfnamefont {S.}~\bibnamefont
  {Venkatapathi}}, \bibinfo {author} {\bibfnamefont {B.}~\bibnamefont {Dong}},
  \ and\ \bibinfo {author} {\bibfnamefont {C.}~\bibnamefont {Hin}},\ }\href
  {\doibase 10.1063/1.4887071} {\bibfield  {journal} {\bibinfo  {journal} {J.
  Appl. Phys.}\ }\textbf {\bibinfo {volume} {116}},\ \bibinfo {pages} {013708}
  (\bibinfo {year} {2014})}\BibitemShut {NoStop}%
\bibitem [{\citenamefont {Bilc}\ \emph
  {et~al.}(2006{\natexlab{a}})\citenamefont {Bilc}, \citenamefont {Mahanti},\
  and\ \citenamefont {Kanatzidis}}]{bilc2006}%
  \BibitemOpen
  \bibfield  {author} {\bibinfo {author} {\bibfnamefont {D.}~\bibnamefont
  {Bilc}}, \bibinfo {author} {\bibfnamefont {S.}~\bibnamefont {Mahanti}}, \
  and\ \bibinfo {author} {\bibfnamefont {M.}~\bibnamefont {Kanatzidis}},\
  }\href {\doibase 10.1103/PhysRevB.74.125202} {\bibfield  {journal} {\bibinfo
  {journal} {Phys. Rev. B}\ }\textbf {\bibinfo {volume} {74}},\ \bibinfo
  {pages} {125202} (\bibinfo {year} {2006}{\natexlab{a}})}\BibitemShut
  {NoStop}%
\bibitem [{\citenamefont {Yamini}\ \emph {et~al.}(2013)\citenamefont {Yamini},
  \citenamefont {Ikeda}, \citenamefont {Pei}, \citenamefont {Doua},\ and\
  \citenamefont {Snyder}}]{Yamini1}%
  \BibitemOpen
  \bibfield  {author} {\bibinfo {author} {\bibfnamefont {S.~A.}\ \bibnamefont
  {Yamini}}, \bibinfo {author} {\bibfnamefont {T.}~\bibnamefont {Ikeda}},
  \bibinfo {author} {\bibfnamefont {A.~L.~Y.}\ \bibnamefont {Pei}}, \bibinfo
  {author} {\bibfnamefont {S.}~\bibnamefont {Doua}}, \ and\ \bibinfo {author}
  {\bibfnamefont {G.~J.}\ \bibnamefont {Snyder}},\ }\href@noop {} {\bibfield
  {journal} {\bibinfo  {journal} {J. Mater. Chem. A}\ }\textbf {\bibinfo
  {volume} {1}},\ \bibinfo {pages} {8725} (\bibinfo {year} {2013})}\BibitemShut
  {NoStop}%
\bibitem [{\citenamefont {Shoenberg}(1984)}]{Shoenberg}%
  \BibitemOpen
  \bibfield  {author} {\bibinfo {author} {\bibfnamefont {D.}~\bibnamefont
  {Shoenberg}},\ }\href@noop {} {\emph {\bibinfo {title} {Magnetic Oscillations
  in Metals}}}\ (\bibinfo  {publisher} {Cambridge University Press},\ \bibinfo
  {year} {1984})\BibitemShut {NoStop}%
\bibitem [{\citenamefont {Martinez}\ \emph {et~al.}(1975)\citenamefont
  {Martinez}, \citenamefont {Schl{\"{u}}ter},\ and\ \citenamefont
  {Cohen}}]{Martinez1}%
  \BibitemOpen
  \bibfield  {author} {\bibinfo {author} {\bibfnamefont {G.}~\bibnamefont
  {Martinez}}, \bibinfo {author} {\bibfnamefont {M.}~\bibnamefont
  {Schl{\"{u}}ter}}, \ and\ \bibinfo {author} {\bibfnamefont {M.~L.}\
  \bibnamefont {Cohen}},\ }\href@noop {} {\bibfield  {journal} {\bibinfo
  {journal} {Phys.\ Rev. B}\ }\textbf {\bibinfo {volume} {11}},\ \bibinfo
  {pages} {651} (\bibinfo {year} {1975})}\BibitemShut {NoStop}%
\bibitem [{\citenamefont {Wiendlocha}(2013)}]{Wiend1}%
  \BibitemOpen
  \bibfield  {author} {\bibinfo {author} {\bibfnamefont {B.}~\bibnamefont
  {Wiendlocha}},\ }\href@noop {} {\bibfield  {journal} {\bibinfo  {journal}
  {Phys. \ Rev. B}\ }\textbf {\bibinfo {volume} {88}},\ \bibinfo {pages}
  {205205} (\bibinfo {year} {2013})}\BibitemShut {NoStop}%
\bibitem [{\citenamefont {Nemov}\ and\ \citenamefont {Ravich}(1998)}]{Nemov1}%
  \BibitemOpen
  \bibfield  {author} {\bibinfo {author} {\bibfnamefont {S.~A.}\ \bibnamefont
  {Nemov}}\ and\ \bibinfo {author} {\bibfnamefont {Y.~I.}\ \bibnamefont
  {Ravich}},\ }\href@noop {} {\bibfield  {journal} {\bibinfo  {journal} {Sov.
  Phys. Usp.}\ }\textbf {\bibinfo {volume} {41}},\ \bibinfo {pages} {735}
  (\bibinfo {year} {1998})}\BibitemShut {NoStop}%
\bibitem [{\citenamefont {Ravich}\ \emph
  {et~al.}(1971{\natexlab{a}})\citenamefont {Ravich}, \citenamefont {Efimova},\
  and\ \citenamefont {Tamarchenkov}}]{Ravich2}%
  \BibitemOpen
  \bibfield  {author} {\bibinfo {author} {\bibfnamefont {Y.~I.}\ \bibnamefont
  {Ravich}}, \bibinfo {author} {\bibfnamefont {A.}~\bibnamefont {Efimova}}, \
  and\ \bibinfo {author} {\bibfnamefont {V.~I.}\ \bibnamefont {Tamarchenkov}},\
  }\href@noop {} {\bibfield  {journal} {\bibinfo  {journal} {Phys. Stat. sol
  (b)}\ }\textbf {\bibinfo {volume} {43}},\ \bibinfo {pages} {11} (\bibinfo
  {year} {1971}{\natexlab{a}})}\BibitemShut {NoStop}%
\bibitem [{\citenamefont {Ravich}\ \emph
  {et~al.}(1971{\natexlab{b}})\citenamefont {Ravich}, \citenamefont {Efimova},\
  and\ \citenamefont {Tamarchenkov}}]{Ravich3}%
  \BibitemOpen
  \bibfield  {author} {\bibinfo {author} {\bibfnamefont {Y.~I.}\ \bibnamefont
  {Ravich}}, \bibinfo {author} {\bibfnamefont {A.}~\bibnamefont {Efimova}}, \
  and\ \bibinfo {author} {\bibfnamefont {V.~I.}\ \bibnamefont {Tamarchenkov}},\
  }\href@noop {} {\bibfield  {journal} {\bibinfo  {journal} {Phys. Stat. sol
  (b)}\ }\textbf {\bibinfo {volume} {43}},\ \bibinfo {pages} {453} (\bibinfo
  {year} {1971}{\natexlab{b}})}\BibitemShut {NoStop}%
\bibitem [{\citenamefont {Cuff}\ \emph {et~al.}(1962)\citenamefont {Cuff},
  \citenamefont {Ellett}, \citenamefont {Kuglin},\ and\ \citenamefont
  {Williams}}]{Cuff1}%
  \BibitemOpen
  \bibfield  {author} {\bibinfo {author} {\bibfnamefont {K.~F.}\ \bibnamefont
  {Cuff}}, \bibinfo {author} {\bibfnamefont {M.~R.}\ \bibnamefont {Ellett}},
  \bibinfo {author} {\bibfnamefont {C.~D.}\ \bibnamefont {Kuglin}}, \ and\
  \bibinfo {author} {\bibfnamefont {L.~R.}\ \bibnamefont {Williams}},\
  }\href@noop {} {\bibfield  {journal} {\bibinfo  {journal} {Proceedings of the
  international conference on the physics of semiconductors in Exeter}\ ,\
  \bibinfo {pages} {316}} (\bibinfo {year} {1962})}\BibitemShut {NoStop}%
\bibitem [{\citenamefont {Kane}(1975)}]{Kanebook}%
  \BibitemOpen
  \bibfield  {author} {\bibinfo {author} {\bibfnamefont {E.~O.}\ \bibnamefont
  {Kane}},\ }\href@noop {} {\emph {\bibinfo {title} {Semiconductors and
  Semimetals, Vol. 1, Chapter 3}}}\ (\bibinfo  {publisher} {Academic, New
  York},\ \bibinfo {year} {1975})\BibitemShut {NoStop}%
\bibitem [{\citenamefont {Ravich}(1970)}]{Ravichbook}%
  \BibitemOpen
  \bibfield  {author} {\bibinfo {author} {\bibfnamefont {Y.~I.}\ \bibnamefont
  {Ravich}},\ }\href@noop {} {\emph {\bibinfo {title} {Semiconducting Lead
  Chalcogenides}}}\ (\bibinfo  {publisher} {Springer},\ \bibinfo {year}
  {1970})\BibitemShut {NoStop}%
\bibitem [{\citenamefont {Bilc}\ \emph
  {et~al.}(2006{\natexlab{b}})\citenamefont {Bilc}, \citenamefont {Mahanti},\
  and\ \citenamefont {Kanatzidis}}]{Bilc1}%
  \BibitemOpen
  \bibfield  {author} {\bibinfo {author} {\bibfnamefont {D.~I.}\ \bibnamefont
  {Bilc}}, \bibinfo {author} {\bibfnamefont {S.~D.}\ \bibnamefont {Mahanti}}, \
  and\ \bibinfo {author} {\bibfnamefont {M.~G.}\ \bibnamefont {Kanatzidis}},\
  }\href@noop {} {\bibfield  {journal} {\bibinfo  {journal} {Phys. Rev. B}\
  }\textbf {\bibinfo {volume} {74}},\ \bibinfo {pages} {125202} (\bibinfo
  {year} {2006}{\natexlab{b}})}\BibitemShut {NoStop}%
\bibitem [{\citenamefont {Kong}(2008)}]{Kong1}%
  \BibitemOpen
  \bibfield  {author} {\bibinfo {author} {\bibfnamefont {H.}~\bibnamefont
  {Kong}},\ }\href@noop {} {\emph {\bibinfo {title} {{Ph.D. Thesis}:
  {Thermoelectric} Property Studies on Lead Chalcogenides, double-filled cobalt
  tri-antimonide and rare earth-ruthenium-germanium}}}\ (\bibinfo  {publisher}
  {University of Michigan},\ \bibinfo {year} {2008})\BibitemShut {NoStop}%
\bibitem [{\citenamefont {Kane}(1956)}]{Kane1}%
  \BibitemOpen
  \bibfield  {author} {\bibinfo {author} {\bibfnamefont {E.~O.}\ \bibnamefont
  {Kane}},\ }\href@noop {} {\bibfield  {journal} {\bibinfo  {journal} {J. Phys.
  Chern. Solids}\ }\textbf {\bibinfo {volume} {1}},\ \bibinfo {pages} {249}
  (\bibinfo {year} {1956})}\BibitemShut {NoStop}%
\bibitem [{\citenamefont {Cohen}(1961)}]{cohen1}%
  \BibitemOpen
  \bibfield  {author} {\bibinfo {author} {\bibfnamefont {M.~H.}\ \bibnamefont
  {Cohen}},\ }\href@noop {} {\bibfield  {journal} {\bibinfo  {journal} {Phys.
  Rev.}\ }\textbf {\bibinfo {volume} {121}},\ \bibinfo {pages} {387} (\bibinfo
  {year} {1961})}\BibitemShut {NoStop}%
\bibitem [{\citenamefont {Dixon}\ and\ \citenamefont {Riedl}(1965)}]{Dixon1}%
  \BibitemOpen
  \bibfield  {author} {\bibinfo {author} {\bibfnamefont {J.~R.}\ \bibnamefont
  {Dixon}}\ and\ \bibinfo {author} {\bibfnamefont {H.~R.}\ \bibnamefont
  {Riedl}},\ }\href@noop {} {\bibfield  {journal} {\bibinfo  {journal} {Phys.
  Rev.}\ }\textbf {\bibinfo {volume} {138}},\ \bibinfo {pages} {A873} (\bibinfo
  {year} {1965})}\BibitemShut {NoStop}%
\bibitem [{\citenamefont {Lin}\ and\ \citenamefont {Kleinman}(1966)}]{Lin1}%
  \BibitemOpen
  \bibfield  {author} {\bibinfo {author} {\bibfnamefont {P.~J.}\ \bibnamefont
  {Lin}}\ and\ \bibinfo {author} {\bibfnamefont {L.}~\bibnamefont {Kleinman}},\
  }\href@noop {} {\bibfield  {journal} {\bibinfo  {journal} {Phys. Rev.}\
  }\textbf {\bibinfo {volume} {142}},\ \bibinfo {pages} {478} (\bibinfo {year}
  {1966})}\BibitemShut {NoStop}%
\bibitem [{\citenamefont {Dimmock}\ and\ \citenamefont
  {Wright}(1964)}]{Dimmock1}%
  \BibitemOpen
  \bibfield  {author} {\bibinfo {author} {\bibfnamefont {J.~O.}\ \bibnamefont
  {Dimmock}}\ and\ \bibinfo {author} {\bibfnamefont {G.~B.}\ \bibnamefont
  {Wright}},\ }\href@noop {} {\bibfield  {journal} {\bibinfo  {journal} {Phys.
  Rev.}\ }\textbf {\bibinfo {volume} {135}},\ \bibinfo {pages} {A821} (\bibinfo
  {year} {1964})}\BibitemShut {NoStop}%
\bibitem [{\citenamefont {Tung}\ and\ \citenamefont {Cohen}(1969)}]{Tung1}%
  \BibitemOpen
  \bibfield  {author} {\bibinfo {author} {\bibfnamefont {Y.~W.}\ \bibnamefont
  {Tung}}\ and\ \bibinfo {author} {\bibfnamefont {M.~L.}\ \bibnamefont
  {Cohen}},\ }\href@noop {} {\bibfield  {journal} {\bibinfo  {journal} {Phys.
  Rev.}\ }\textbf {\bibinfo {volume} {180}},\ \bibinfo {pages} {823} (\bibinfo
  {year} {1969})}\BibitemShut {NoStop}%
\end{thebibliography}

%%%%%%%%%%%%%%%%%%%%%%% BIBLIOGRAPHY


%%%%%%%%%%%%%%%%%%%%%%%%%%%%%%%%%%%%%%%%%

\vspace{2cm}

\appendix
\section*{Appendix}
\setcounter{figure}{0} 
\makeatletter 
\renewcommand{\thefigure}{A\@arabic\c@figure}
\makeatother

\section{Quantum oscillations formalism}\label{app_QO}

In this section we briefly outline the concepts needed to understand quantum oscillation experiments in metals. For a detailed treatment see the excellent book by \citet{Shoenberg}. 
It is well known that in a magnetic field $H$ the allowed electronic states lie on quantized tubes in $k$-space (Landau tubes). The tube quantization is described by the Onsager equation
\begin{equation}
	a(E_{n,k_H},k_H)=\left(n+\frac{1}{2}\right)2\pi eH/\hbar c \quad ,
\end{equation}
where $a$ is the cross-sectional area of the Landau tube in a plane perpendicular to $H$, and $n$ is an integer. As a consequence, an oscillatory behavior with the inverse magnetic field $1/H$ can be observed in, for example the magnetization -- the de Haas-van Alphen (dHvA) effect -- or the resistance -- the Shubnikov-de Haas effect. 
The period of such oscillations, $\Delta_{1/H}$, is given by
\begin{equation}
	\Delta_{1/H}=2\pi e/(\hbar c A) \quad ,
\end{equation}
$A$ being an \textit{extremal} cross-sectional area of the Fermi surface in a plane perpendicular to $H$. 
One can also define a frequency for these oscillations as
\begin{equation}
	f=1/\Delta_{1/H}=(c\hbar/2\pi e)A \quad .
\end{equation}
By determining the oscillations in, e.g., the resistivity for varying orientations of the magnetic field one can eventually reconstruct the Fermi surface. 

In the semi-classical picture the electrons move along (open and closed) orbits on the Fermi surface in a plane perpendicular to H. The time taken to traverse a closed (cyclotron) orbit is given by
\begin{equation}
	t_c=\frac{2\pi}{\omega_c}=\frac{\hbar^2c}{eH}\frac{\partial a}{\partial E} \quad ,
\end{equation}
where one can rewrite the cyclotron frequency $\omega_c$ in terms of a cyclotron mass
\begin{equation}
\label{eq_cyc_masses}
	m^\mathrm{cyc}=\frac{\hbar^2}{2\pi}\frac{\partial a}{\partial E}\,.
\end{equation}
For a free-electron gas the cyclotron mass is equal to the electron mass. 
Experimentally the cyclotron masses are extracted using the Lifshitz-Kosevich (LK) formula (in SI units)

\begin{eqnarray}
\frac{\rho(H) -\rho_{0}}{\rho_0} & & =\sum_{i} C_i \left\{\exp\left(\frac{-\num{14.7}(m_i^{cyc}/m_0) \Theta_{D,i})}{H}\right)\right\} \nonumber \\
& & \times\left\{\frac{T/H}{\sinh\left(\num{14.7}(m_i^{cyc}/m_0)T/H\right)}\right\} \nonumber \\
& & \times\cos\left[2\pi\frac{f_i}{H}+\phi_i\right] 
\end{eqnarray}

as presented in Eqn. \ref{eq_LKmass}.

\section{Cyclotron effective mass anisotropy} \label{app_massan}

For a given dispersion relation one can, in principle, find the relation between the geometric anisotropy of the Fermi surface and the anisotropy of the cyclotron effective mass. For a perfect parabolic band, the general anisotropic dispersion relation is given by,

\begin{equation}
\frac{\hbar^2k_x^2}{2m_x}+\frac{\hbar^2k_y^2}{2m_y}+\frac{\hbar^2k_z^2}{2m_z}=E \label{disppar}
\end{equation} 

where $m_x$, $m_y$ and $m_z$ are the band masses. For an ellipsoidal Fermi surface with the major semiaxis of the ellipse oriented along the $z$-axis, the band masses are $m_x=m_y=m_{\bot}$ and $m_z=m_{\|}$. For such systems, the minimum and maximum cross sectional areas are

\begin{subequations}
\begin{align}
A_{\bot}&=\pi k_{x,y}^2\bigg|_{k_z=0}=\frac{2\pi m_{\bot}}{\hbar^2}E\\
A_{\|}&=\pi k_{x,y}\bigg|_{k_{z}=0} k_{z}\bigg|_{k_{x,y}=0}=\frac{2\pi}{\hbar^2}\sqrt{m_{\bot}m_{\|}}E
\end{align}
\end{subequations}

and the ratio of maximum-to-minimum cross sectional areas is

\begin{equation}
\frac{A_{\|}}{A_{\bot}}=\sqrt{\frac{m_{\|}}{m_{\bot}}}=\sqrt{K}
\end{equation}

where $K=m_{\|}/m_{\bot}$ is defined as the ratio of band masses, and it directly represents the anisotropy of the ellipsoidal pocket. 
As our experiment is a direct probe of cyclotron masses, we can find a relation between $K$ and the extremal cyclotron masses. For a perfect parabolic band, with dispersion of the form given in Eqn. \ref{disppar}, the cyclotron effective mass, $m^{cyc}=e|\vec{B}|/\hbar\omega_c$, for a magnetic field of the general form $\vec{B}=B_x\hat{x}+B_y\hat{y}+B_z\hat{z}$ can be found from the dynamic equations and the dispersion relation, resulting in the expression

\begin{equation}
m^{cyc}=\sqrt{\frac{m_x m_y m_z}{m_x\left(\frac{B_x}{|B|}\right)^2+m_y\left(\frac{B_y}{|B|}\right)^2+m_z\left(\frac{B_z}{|B|}\right)^2}} \label{masscyc}
\end{equation}

For an ellipsoid of revolution, therefore, the transverse and longitudinal cyclotron effective masses, in terms of the band masses, are

\begin{subequations}
\begin{align}
m^{cyc}_{\bot}&=m_{\bot}\\
m^{cyc}_{\|}&=\sqrt{m_{\bot}m_{\|}}
\end{align}
\end{subequations} 

And with this, 

\begin{equation}
\frac{m^{cyc}_{\|}}{m^{cyc}_{\bot}}=\sqrt{\frac{m_{\|}}{m_{\bot}}}=\sqrt{K}
\end{equation}

For the case of Pb$_{1-x}$Na$_x$Te, in which $K=$14.3$\pm$0.4 for a wide range of dopings, $m^{cyc}_{\|}/m^{cyc}_{\bot}=$3.78$\pm$0.05. Additionally, from Eqn. \ref{masscyc}, we can find a general expression for the angle dependence of cyclotron mass of an ellipsoid of revolution, with respect to the main axis of the ellipsoid, and as a function of the transverse cyclotron mass, by writing the components of the magnetic field in spherical coordinates as $B_x=|\vec{B}|\sin{\theta}\cos{\varphi}$, $B_y=|\vec{B}|\sin{\theta}\sin{\varphi}$ and $B_z=|\vec{B}|\cos{\theta}$:

\begin{equation}
\frac{m^{cyc}\left({\theta}\right)}{m^{cyc}_{\bot}}=\sqrt{\frac{K}{(K-1)\cos^2\theta+1}} \label{masscycangle}
\end{equation}

\section{Effective cyclotron mass along the [100] orientation}
 
\begin{figure}[H]
%\vspace{-0.3cm}
\hspace{-0.2cm}
\centering
\includegraphics[scale=0.6]{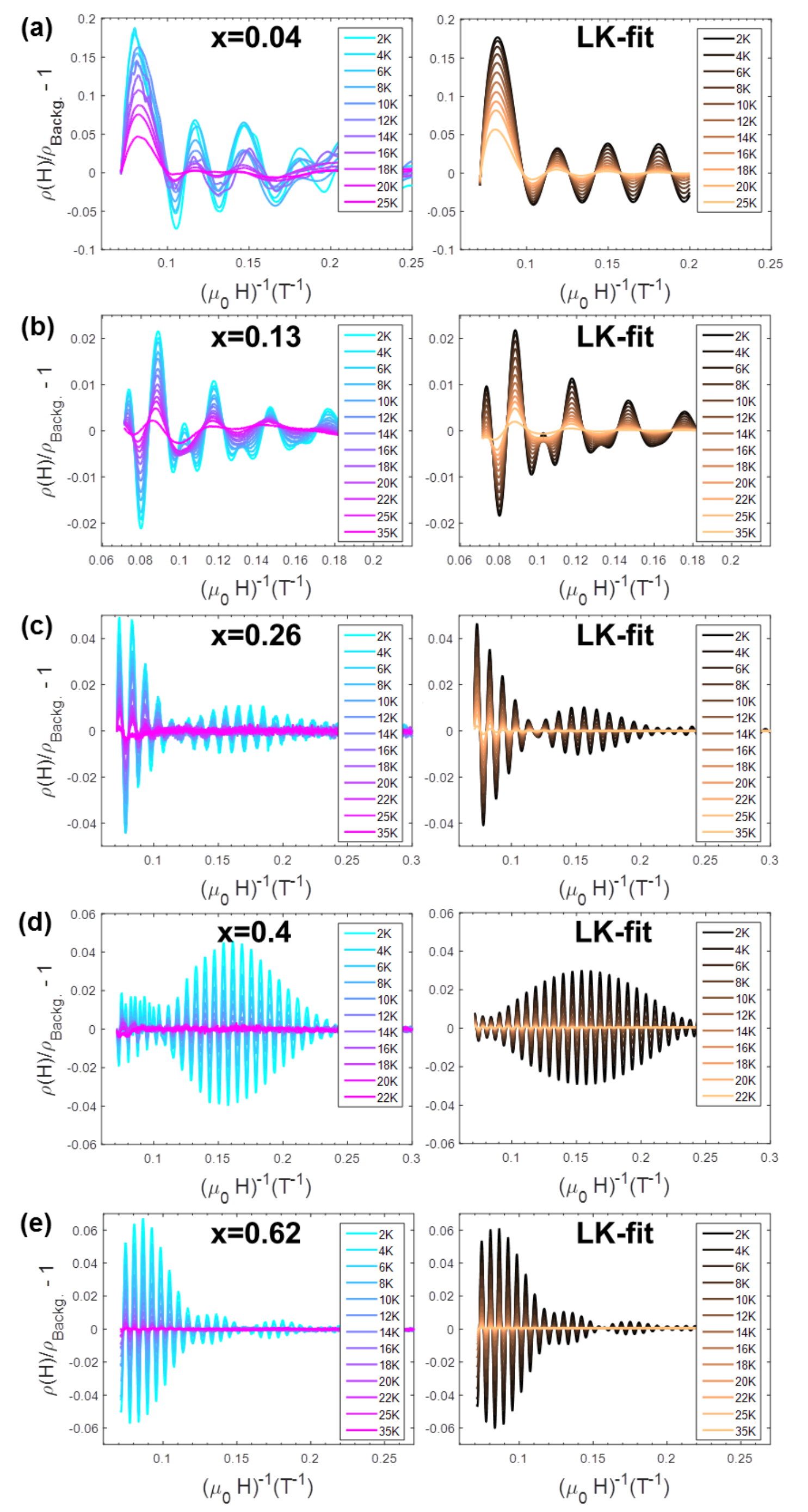}
\caption{(Color online) Temperature dependence of the amplitude of the oscillating component of magnetoresistance for Pb$_{1-x}$Na$_x$Te samples, with magnetic field oriented in or close to the [100] direction. The left-column plots of each composition show the background-free data at different temperatures. The right-column plots show the fits of the data to the LK-formula in equation \ref{eq_LKmass}, using the two most dominant frequencies observed in the FFT of the lowest temperature curve (three most dominant for the $x$=0.62$\%$ sample). From these fits, the values of cyclotron effective mass and Dingle temperature, for each frequency term, are obtained.
}\label{fig_Tdepall100}
%\vspace{-0.4cm}
\end{figure}

Figure \ref{fig_Tdepall100} shows the temperature dependence of the oscillating component of magnetoresistance for Pb$_{1-x}$Na$_x$Te samples of different Na concentrations, for field oriented along the [100] direction, which provides direct access to the $m_{[100]}^{cyc}$ cyclotron effective mass. Least-squares fits to equation \ref{eq_LKmass}, including up to the second strongest frequency component, for each Na doping, and for a field range of 3-5 T to 14 T, are shown in the right-column plots of this figure. The obtained [100] cyclotron masses are summarized in Table \ref{table_mass}, and plotted as a function of carrier concentration in Fig. \ref{fig_massAll}, in the discussion section.

\end{document}